\documentclass[a4paper,11pt]{article}

\pdfoutput=1 
             
\usepackage{draftwatermark}

\usepackage{jcappub} 
\usepackage{lineno}

\usepackage[T1]{fontenc} 

\title{Exploring cosmic homogeneity with the BOSS DR12 galaxy sample}

\author[a,1]{Pierros Ntelis,}
\author[a]{Jean-Christophe Hamilton,}
\author[b]{Jean-Marc Le Goff,}
\author[b]{Etienne Burtin,}
\author[b]{Pierre Laurent,}
\author[b]{James Rich,}
\author[a]{Nicolas Guillermo Busca,}
\author[m]{Jeremy Tinker,}
\author[a]{Eric Aubourg,}
\author[b]{H\'elion du Mas des Bourboux,}
\author[f]{Julian Bautista,}
\author[b]{Nathalie Palanque Delabrouille,}
\author[g]{Timoth\'ee Delubac,} 
\author[d]{Sarah Eftekharzadeh,}
\author[c]{David W. Hogg,} 
\author[d]{Adam Myers,}
\author[j]{Mariana Vargas-Maga\~na,}
\author[e]{Isabelle P\^aris,}
\author[l]{Partick Petitjean,}
\author[h]{Graziano Rossi,}
\author[k]{Donald P. Schneider,}
\author[i]{Rita Tojeiro}
\author[b]{and Christophe Yeche}

\affiliation[a]{APC, Universit\'{e} Paris Diderot-Paris 7, CNRS/IN2P3, CEA, Observatoire de Paris,\\ 10, rue A. Domon \& L. Duquet,  Paris, France}
\affiliation[b]{IRFU, CEA, Universit\'{e} Paris-Saclay, F-91191, Gif-sur-Yvette, France}
\affiliation[c]{Center for Cosmology and Particle Physics, New York University, 4 Washington Place, Meyer Hall of Physics, New York, NY 10003, USA}
\affiliation[d]{Department of Physics and Astronomy, University of Wyoming, Laramie, WY 82071, USA}
\affiliation[e]{Aix Marseille Universite, CNRS, LAM (Laboratoire d?Astrophysique de Marseille) UMR 7326,
13388, Marseille, France}
\affiliation[f]{Department of Physics and Astronomy, University of Utah, Salt Lake City, UT 84112, USA.}
\affiliation[g]{Laboratoire d'astrophysique, Ecole Polytechnique F\'ed\'erale de Lausanne (EPFL), Observatoire de Sauverny,CH-1290 Versoix, Switzerland}
\affiliation[h]{Department of Astronomy and Space Science, Sejong University, Seoul, 143-747, Korea}
\affiliation[i]{School of Physics and Astronomy, University of St Andrews, St Andrews, KY16 9SS, UK}
\affiliation[j]{Instituto de F\'{i}sica, Universidad Nacional Aut\'{o}noma de M\'{e}xico, Apdo. Postal 20-364, M\'{e}xico}
\affiliation[k]{Department of Astronomy and Astrophysics, The Pennsylvania State University, University Park, PA 16802}
\affiliation[l]{Institut d?Astrophysique de Paris, CNRS-UPMC, UMR7095, 98bis bd Arago, Paris, 75014 France}
\affiliation[m]{Department of Physics and Center for Cosmology and Particle Physics, New York University}

\emailAdd{pntelis <at> apc <point> in2p3 <point> fr}

\abstract{
In this study, we probe the transition to cosmic homogeneity in the Large Scale Structure (LSS) of the Universe using the CMASS galaxy sample of BOSS spectroscopic survey which covers the largest effective volume to date, $3\ h^{-3}\ \mathrm{Gpc}^3$ at $0.43 \leq z \leq 0.7$. We study the scaled counts-in-spheres, $\mathcal{N}(<r)$, and the fractal correlation dimension, $\mathcal{D}_2(r)$, to assess the homogeneity scale of the universe using a $Landy\ \&\ Szalay$ inspired estimator. 

Defining the scale of transition to homogeneity as the scale at which $\mathcal{D}_2(r)$ reaches 3 within $1\%$, i.e. $\mathcal{D}_2(r)>2.97$ for $r>\mathcal{R}_H$, we find $\mathcal{R}_H = (63.3\pm0.7) \ h^{-1}\ \mathrm{Mpc}$, in agreement at the percentage level with the predictions of the $\Lambda$CDM model $\mathcal{R}_H=62.0\ h^{-1}\ \mathrm{Mpc}$. Thanks to the large cosmic depth of the survey, we investigate the redshift evolution of the transition to homogeneity scale and find agreement with the $\Lambda$CDM prediction. Finally, we find that $\mathcal{D}_2$ is compatible with $3$ at scales larger than $300\ h^{-1}\ $Mpc in all redshift bins.

These results consolidate the Cosmological Principle and represent a precise consistency test of the $\Lambda CDM$ model.}

\begin{document}
\maketitle
\flushbottom

\section{Introduction}
Most models of modern Cosmology are based on solutions of General Relativity for an isotropic and homogeneous universe. The standard model, known as $\Lambda CDM$, is mainly composed of a Cold Dark Matter (CDM) and $\Lambda$ corresponds to a cosmological constant. 
This model shows excellent agreement with current data, be it from Type Ia supernovae~\cite{SnIaPerlmutter,SnIaRiess}, temperature and polarisation anisotropies in the Cosmic Microwave Background~\cite{CMB-Planck} or Large Scale Structure~\cite{LssPercival,LssParkinsonWg,LssHeymans2013Cfhtlens}. The two main assumptions of this model are the validity of General Relativity and the {\em Cosmological Principle}~\cite{CP} that states that the Universe is isotropic
and homogeneous on large enough scales, or equivalently that the Universe statistical properties are both rotationally and  translationally invariant on large scales. 
	
Isotropy is well tested through several probes at various cosmic epochs: at $z\approx1100$, Cosmic Microwave Background temperature anisotropies, corresponding to density fluctuations in the young Universe, have been shown to be of order $10^{-5}$~\cite{CMB-Cobe}. In the more recent Universe, the distribution of sources in X-ray~\cite{X-ray-ref1} and radio~\cite{Radio} surveys strongly supports isotropy. 
Large spectroscopic galaxy surveys, such as the baryon oscillation spectroscopic survey (BOSS) of the third Sloan digital sky survey (SDSS-III), show no evidence for anisotropies in the projected galaxy distribution in volumes of a few $\mathrm{Gpc}^3$~\cite{boss2011}.

We should stress, however, that these observations test isotropy at a given redshift or after projection over a given range of redshifts. This "projected'' isotropy is a weaker assumption than ``spatial'' isotropy, which is an isotropy at each redshift~\footnote{More precisely, spatial isotropy is the assumption that $\rho(r,\theta_{1})=\rho(r,\theta_{2})$ for any given $(r,\theta_{1},\theta_{2})$, while projected isotropy is the assumption that $\rho(\theta_{1})=\rho(\theta_{2})$ for any $(\theta_{1},\theta_{2})$, where $\rho(\theta) = \int \rho(r,\theta) W(r) dr$ and $W(r)$ is the window function. }.
Combining spatial isotropy and the Copernican principle, which states that our position in the Universe is not privileged, implies that the Universe is homogeneous~\cite{CP, Maartens,clarkson2012establishing,straumann1974minimal}. However, as shown by Durrer et al.~\cite{durrer1997angular}, this implication is not true if we only have projected isotropy. So CMB isotropy, for instance, cannot be combined with the Copernican principle to prove homogeneity.

It is therefore important to test homogeneity. Large three dimensional spectroscopic surveys offer an excellent occasion to strengthen this aspect of the cosmological model with accurate observations. Most of the studies conducted so far found a transition to homogeneity in the galaxy distributions at scale between 70 and $150\ h^{-1}$ Mpc~\cite{Hogg,Yadav,Yadav1,Sarkar,Martinez,Guzzo,Martinez2,Scaramella,Amendola,PanColes,WiggleZ,sarkar2016many} therefore strengthening the cosmological principle.  Some studies did not find a transition to homogeneity~\cite{Appleby,Sylos,Pietronero,Labini2,Joyce,Labini3,Labini4}. However none of those studies reached scale larger than 200 $h^{-1}$ Mpc, where homogeneity becomes clear.

However, these 3D surveys investigate the statistical properties on the observer past light-cone and not inside of it, so they do not observe galaxies at the same epoch. An attempt to overcome this limitation was to use star formation history in order to study the homogeneity inside the past light-cone~\cite{hoyle2012testing}, but this is model dependent. Another possibility~\cite{zibin2014nowhere} is to use a combination of secondary CMB probes, including integrated Sachs-Wolfe, kinetic Sunyaev-Zel'dovich and Rees-Sciama effects.
	
In this article, we use the Data Release 12 of the SDSS-III/BOSS spectroscopic galaxy sample to search for a transition to homogeneity~\cite{wu1999large}. The BOSS CMASS galaxy catalogue covers a volume of $5.1$ $h^{-3}$ Gpc$^3$ in the redshift range $0.43 < z < 0.7$ and contains nearly one million galaxies in 8500 square degrees. It represents the largest effective volume sampled as of today ~\footnote{ The effective volume of a survey is given by $V_{\rm eff} = V \frac{nP}{1 + nP} $, which describes the statistical power of the sample. The CMASS sample has $V^{\rm CMASS}_{\rm eff} \simeq 2.9h^{-3}\mathrm{Gpc}^{3}$, while the QSO sample has $V^{\rm QSO}_{\rm eff} \simeq 0.21h^{-3}\mathrm{Gpc}^{3}$.
} 
and therefore offers a unique opportunity to observe the transition to homogeneity. A similar study of homogeneity has been performed with the DR12 quasar sample of BOSS \cite{Laurent}. We mostly follow the method introduced in~\cite{WiggleZ} to measure the "fractal correlation dimension'' of the distribution, $\mathcal{D}_2(r)$, as an indicator of the transition to homogeneity\cite{wu1999large}. 

Due to the possible evolution of the tracer with redshift, a redshift survey is blind to strictly radial variation of the density~\cite{Mustapha}.
So strictly speaking we can only test homogeneity up to a radial variation, that is we can test spatial isotropy (and then combining spatial isotropy with the Copernican principle we can prove homogeneity). On the other hand, as discussed by \citet{Laurent}, we can demonstrate spatial isotropy independently of a fiducial cosmology~\footnote{
Indeed let's assume a homogeneous 3D spectroscopic survey with a constant density $\rho(r,\theta,\phi)=\mathrm{Cst}$. This density is the product of the Jacobian $J(z) = H(z)/cD^2(z)$ with the observed density of sources $dN(\theta,\phi,z)/d\Omega dz$ which is a function of angles and redshift. As a result the observed density is necessarily a function of redshift only following: $dN(\theta,\phi,z)/d\Omega dz \propto J^{-1}(z)$, which is the definition of spatial isotropy.
}.

Throughout this study, unless stated otherwise, we use a fiducial flat $\Lambda$CDM cosmological model with the following parameters:
\begin{equation}\label{fid-cosmo}
		\textbf{p}_{\rm cosmo}=(\omega_{cdm},\omega_{b},h,n_{s},\ln\left[10^{10}A_{s}\right]) = (0.1198,0.02225,0.6727,0.9645,3.094) \; ,
\end{equation}
where $\omega_{b} = \Omega_{b}h^{2}$  and $\omega_{\rm cdm} = \Omega_{\rm cdm}h^{2}$ are the reduced fractional density of baryons and  cold dark matter, respectively;   $h=H_0/[100$ km s$^{-1}$ Mpc$^{-1}]$, with $H_0$ the  Hubble constant;
and finally  $n_{s}$ and $A_{s}$ are, respectively,  the spectral index and the amplitude of the primordial scalar power spectrum.
The numerical values in Eq.~\ref{fid-cosmo} are from Planck $2015\ TT,TE,EE+lowP$ analysis~\cite{Params-Planck}. 
	
	The paper is organized as follows. Section 2 describes the galaxy data sample used in this analysis and section 3  the methodology to quantify cosmic homogeneity. In section 4 we introduce the model that is adjusted to the observed data, while in section 5 we present our results. In section 6 we test the robustness of our method, and we conclude in section 7.

\section{Dataset}
	\subsection{The BOSS survey}

BOSS\cite{dawson2012baryon} is dedicated to studying the 3D distribution of $\sim1.4\times10^{6}$ galaxies and $\sim10^{5}$ quasars and their Lyman-$\alpha$ forests within an effective area of $\sim10,000$ deg$^{2}$ \cite{BOSS-DR12,eisenstein2011sdss}. 
Aluminum plates are set on the focal plane of the $2.5\mathrm{m}$ telescope\cite{Telescope} and drilled with $1000$ holes corresponding to targets to be observed. An optical fiber is fixed on each hole, allowing $1000$ spectra to be measured per exposure~\cite{spectra} with a typical redshift uncertainty of a few tens of $\mathrm{km\cdot s^{-1}}$~\cite{reso_z}. Several color cuts are used to select massive galaxy in the redshift range $0.43<z<0.7$ in the imaging data from SDSS-I-II and SDSS-III/BOSS in $(u,g,r,i,z)$ bands~\cite{imaging}. These cuts are designed to result in a stellar mass limit that is constant with redshift, according to a passively evolving model \cite{Maraston}. The large masses ensure a strong bias with respect to the underlying dark matter field, providing a high signal-to-noise ratio on the two-point correlation function. The resulting sample of galaxies is called CMASS and consists in $\sim 10^6$ galaxies with accurate spectroscopic redshifts. The CMASS sample has been used to measure the Baryonic Acoustic Oscillation feature (BAO) with unprecedented accuracy, putting strong constraints on cosmological parameters~\cite{BAOcosmo}. We use the same galaxy sample to perform a measurement of the transition to homogeneity benefiting from the signal-to-noise enhancement due to the large bias of this galaxy sample.
	
We weight galaxies with~\citep{Target-Selection}:
	\begin{equation}\label{eq:weights}
		w_{gal} = (w_{cp} + w_{noz} - 1) \times w_{star} \times w_{see} \times w_{FKP} \; ,
	\end{equation}
Here, the close-pair weight, $w_{cp}$, accounts for the fact that, due to fiber coating, one cannot assign optical fibers on the same plate to two targets that are closer than $62''$. The  $w_{noz}$ weight accounts for targets for which the pipeline failed to measure the redshift. The $w_{star}$ and $w_{see}$ weights  correct for the dependance of the observed galaxy number density  with the stellar density and with seeing, respectively. Finally, we use the FKP weight, $w_{FKP}$,~\cite{FKP} in order to reduce the variance of the two-point correlation function estimator.

	\begin{table}[h!]
		\begin{center} 
		\begin{tabular}{ *3c } 
	 	$z$   & NGC & SGC  \\ 
	 	\hline 
		0.430-0.484 & 101,383  & 40,170   \\ 
	 	0.484-0.538 & 174,468  & 63,518   \\ 
	 	0.538-0.592 & 151,084  & 56,805   \\ 
		0.592-0.646 &\ 97,155    & 37,179   \\ 
		0.646-0.700 &\ 47,289    & 17,899   \\ 
	 	\hline
		0.430-0.700 & 571,379  & 215,571 \\ 
	 	\hline
		\end{tabular}
		\end{center} 
	\caption{\label{tab:zbins} DR12 data sample in 5 redshift intervals in north (NGC) and south (SGC) galactic caps.}
	\end{table}	
	
\subsection{Data sample}

\begin{figure}[ht!]
	\centering
	\includegraphics[width=1\linewidth, keepaspectratio]{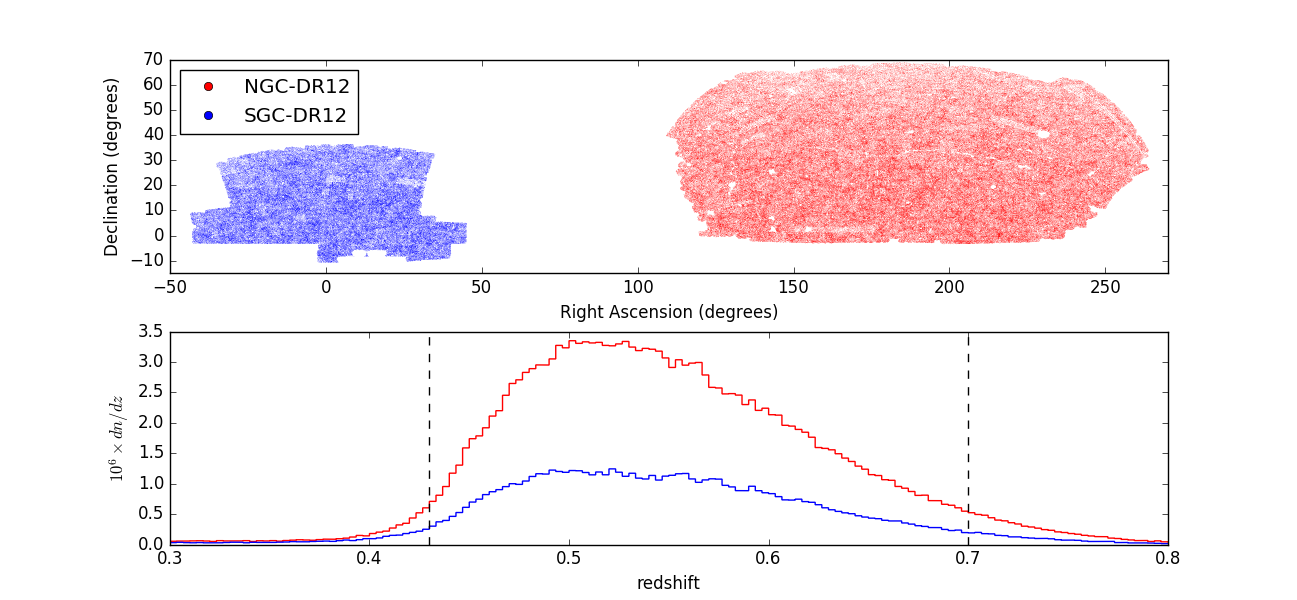}
	\caption{\label{fig:DR12} Right ascension and declination profiles (top) and redshift distribution (bottom) of the DR12 sample for the North (NGC) and South (SGC) Galactic Caps. The vertical dashed lines define the redshift interval used in our analysis.}
	\end{figure}

We divide our data sample into 5 redshift intervals, as defined in table \ref{tab:zbins}, to study the evolution of the clustering of the CMASS galaxy sample. The angular and redshift distributions of the sample are shown in figure \ref{fig:DR12}. We use a flat $\Lambda$CDM model with parameters defined by equation \ref{fid-cosmo} to convert the redshift measurements to comoving distances:
	\begin{equation}\label{eq:FRW}
		d_{\rm comov} (z) =\frac{c}{H_{0}} \int_{0}^{z} \frac{dz'}{E(z')} \;,
	\end{equation}
where $E(z) = \sqrt{\Omega_{m}(1+z)^{3} + \Omega_{\Lambda} }$. A final sample with a total effective volume of 3.8 $h^{-3}$ Gpc$^{3}$ \cite{Target-Selection} is obtained, which is significantly larger than the effective volume used in previous studies, such as 0.6 $h^{-3}$Gpc$^3$ for WiggleZ~\cite{WiggleZ}, $1 h^{-3}Gpc^{3}$ for a DR7 LRG galaxy sample study~\cite{sarkar2016many}  or 0.2 $h^{-3}$Gpc$^{3}$ for the SDSS II LRG analysis~\cite{Hogg} .  

\subsection{Mock-Catalogues}

Mock catalogues are an important tool for determining uncertainties in galaxy surveys and tuning data analysis pipelines. In our analysis we use 1000 mock catalogues constructed with the quick-particle-mesh (QPM) algorithm~\cite{QPM} for BOSS. These are expected to be more realistic than mocks based on second-order Lagrangian perturbation theory, although not as much as the highly time-consuming N-body  simulations. These mock catalogues use a different flat $\Lambda$CDM cosmology than the one we use in our analysis, but this difference is accounted for in our analysis:
	\begin{equation}\label{eq:qpmcosmo}
		\textbf{p}_{\rm{qpm}} = (\Omega_{cdm},\Omega_{b},h,n_{s},\sigma_{8}) = (0.274,0.046,0.7,0.95,0.8)
	\end{equation}
	where $\sigma_{8}$ is the variance of the matter power spectrum computed in spheres of 8 $h^{-1}$ Mpc radius.

\section{Methodology\label{sect:methodology}}

In this section we present the method we have used to study the transition to homogeneity in the distribution of the BOSS CMASS galaxy catalogue. We follow~\citet{WiggleZ} in assessing the homogeneity of the catalogue through the fractal dimension, $D_{2}(r)$, but use significantly different methods.

\subsection{Observables}

The ``counts-in-spheres'', $N(<r)$, is defined as the average number of objects in a sphere of radius $r$, it should obviously scales as $r^3$ for an homogeneous distribution. This scaling is the basis for the definition of the fractal dimension:
	\begin{equation}
		D_{2}(r) \equiv \frac{d\ \ln N(<r)}{d\ \ln r} \; .
	\end{equation}
For a homogeneous distribution one gets $D_{2}(r) = 3$, while for a fractal distribution $D_2 < 3$ is a measure of the fractal dimension. These simple expressions are no longer valid in the case of a survey that has a peculiar geometry and a non-uniform completeness. We correct for these geometrical effects by using catalogues of random points that uniformly fill the survey volume. We use random samples that are five times larger than the data sample in order to ensure that statistical fluctuations due to the random points are significantly smaller than those due to the data. We therefore define a ``scaled counts-in-spheres''
	\begin{equation}
		\mathcal{N}(<r) = \frac{N_{gal} (<r)}{N_{rand} (<r)} \label{eq:defN}
	\end{equation}
 which is the ratio between the counts-in-spheres of the galaxy distribution, $N_{gal} (<r)$, and the counts-in-spheres of the random point distribution inside the same survey geometry, $N_{rand}(<r)$. This quantity is now expected to be independent of $r$ (ratio of two quantities $\propto r^3$) for a homogeneous distribution and therefore the fractal dimension  is redefined as:
	 \begin{equation}\label{eq:Fractal}
	 	\mathcal{D}_{2}(r) \equiv \frac{d\ \ln\ \mathcal{N}(<r)}{d\ \ln\ r} + 3 \ ,
	 \end{equation}
	 in order to be equal to 3 for a homogeneous distribution.
	 
	 $\mathcal{N}(<r)$ and $\mathcal{D}_{2}(r)$ are the two observables we will consider in the analysis, and we show in the next section that there are different ways to obtain them from the data.	
	 	 
 \subsection{Estimators}\label{subsect:Estimators}

We define the following quantities from our data and random catalogues:
	 \begin{itemize}
	 \item  $\displaystyle dd(r) = \frac{DD(r)}{n_{g}(n_{g}-1)/2}$: the normalized number of galaxy pairs distant by $r$,
	 \item $\displaystyle rr(r) = \frac{RR(r)}{n_{r}(n_{r}-1)/2}$: the normalized number of random-point pairs distant by $r$,
	 \item $\displaystyle dr(r) = \frac{DR(r)}{n_{g}n_{r}}$: the normalized number of galaxy random-point pairs distant by $r$,
	 \end{itemize}
	 where $n_{g}$ and  $n_{r}$ are the total number of galaxies and random points, respectively.
The definition of $\mathcal{N}(<r)$ in Eq.~\ref{eq:defN} recalls the Peebles-Hauser estimator, $\widehat \xi(r)=dd(r)/rr(r)-1$, for the two-point correlation function~\cite{PH}. This estimator is known to be less efficient 
than the more sophisticated Landy-Szalay estimator~\cite{LSestimator}. 
	\begin{equation}
		\widehat{\xi}_{ls}(r) = \frac{dd(r) -2dr(r) + rr(r)}{rr(r)} \; ,
	\end{equation}
which has minimal variance on scales where $\xi(r)<<1$\cite{LSestimator}. 

\citet{Laurent} defines a Landy-Szalay inspired estimator for the counts-in-spheres:
	\begin{equation}\label{eq:lau}
		\widehat{\mathcal{N}}_{lau}(<r) = 1+ \frac{\int_{0}^{r}( dd(s) -2dr(s) + rr(s) )ds}{\int_{0}^{r}rr(s)ds}\ .
	\end{equation}
Alternatively, we can directly compute the counts-in-spheres from the  Landy-Szalay estimator of the two-point correlation function itself,  as explained in appendix \ref{APP:Nr_of_xi}:
	\begin{equation}\label{eq:Counts}
		\widehat{\mathcal{N}}_{cor}(<r) = 1+ \frac{3}{r^{3}} \int_{0}^{r}\widehat{\xi}_{ls}(s)s^{2}ds\ .
	\end{equation}
These two estimators are expected to be closer to optimal than the estimator of equation \ref{eq:defN}.
They result in two different estimators for the fractal dimension that we name accordingly $\widehat{\mathcal{D}}_{2,lau}(r)$ and $\widehat{\mathcal{D}}_{2,cor}(r)$. In appendix \ref{subsec:Choice}, we show that the $cor$-estimators, $\widehat{\mathcal{N}}_{cor}(<r)$ and $\widehat{\mathcal{D}}_{2,cor}(r)$, are less biased than the $lau$-estimators. In our final analysis, unless stated otherwise, we use the $cor$-estimators and we drop the $lau$ indice and the hat for sake of simplicity.

Since we are interested in matter distribution, we must use the galaxy bias, $b$, to convert the counts-in-spheres measured for galaxies to the one we would get if we were measuring the whole matter (mostly dark matter).  We also have to account for redshift space distortions and we explain in details how we do that in section \ref{subsec:RSD}.

\subsection{Homogeneity scale definition}

Following~\citet{WiggleZ}, we define an homogeneity scale $R_{H}$ for both $\mathcal N$ and $D_2$ observables, as the scale for which the observable approaches its homogeneous value within 1\%. We then have two a-priori different transition-to-homogeneity scales defined by:
\begin{equation}\label{eq:RH-definition}
\mathcal{D}_{2}(R^{\mathcal{D}_2=2.97}_H) = 2.97  \quad {\rm and} \quad \mathcal{N}(R^{\mathcal{N}=1.01}_{H}) = 1.01 \; .
\end{equation}
These are, of course, arbitrary definitions for the homogeneity scales but they do not depend on the survey and can be used to test cosmological models and compare different survey measurements as long as the same definitions are used in all cases~\cite{WiggleZ}. 
	
Other authors have proposed different ways of defining the homogeneity scale. In particular~\citet{Yadav1} proposed to define it as the scale at which $\mathcal{D}_{2}$ cannot be distinguished from 3 within the survey errors. This is obviously another proper definition because it is independent of the bias and similar conclusions can be drawn as we demonstrate at the end of section \ref{subsect:ConD2}. Furthermore, using this estimator on the mock catalogues (See section \ref{subsect:ConD2} ), we find larger variations on the estimated homogeneity scale in respect with the one obtained by Eq.~\ref{eq:RH-definition}. For these reasons, we have made the choice in this study to use the arbitrary but universal definition in Eq.~\ref{eq:RH-definition}, although we show that beyond 300 $h^{-1}$Mpc, we do find consistency with $\mathcal{D}_{2}=3$ within our measurement errors.

\section{Theoretical Model}\label{sec:Theoretical Model}
\subsection{Prediction for $\xi(r),\mathcal{N}(<r)$ and $\mathcal{D}_2(r)$}
	
We use the CLASS software \cite{CLASS} to compute the $\Lambda$CDM theoretical prediction for the two-points correlation function and the observables we consider in the analysis. CLASS computes the theoretical matter power spectrum,  $P^{(r)}_{\delta\delta}(k)$, where the $(r)$ exponent indicates that it is calculated in 
real space, by opposition to redshift space.
In order to get a prediction for the  redshift-space galaxy power-spectrum, $P^{(s)}_{gg}(k)$ where the $(s)$ exponent indicates redshift space, we model redshift-space-distortion (RSD):

\begin{itemize}
\item On large scales, galaxies are falling into large gravitational potentials, which tends to sharpen their distribution along the line-of-sight in redshift space. This is known as the Kaiser effect~\cite{Kaiser}, which results in:
	\begin{equation}\label{eq:largeRDS}
		P^{(s)}_{gg}(k,\mu;b) = b^{2}  (1+\beta \mu^{2})^{2} \ P^{(r)}_{\delta\delta}(k) \; ,
	\end{equation}
where $b$ is the linear (scale independent)  bias between galaxy and matter distributions, $\beta = f/b$ where $f$ is the linear growth rate, which can be approximated by $f \approx \Omega^{0.55}_{m}(z)$\cite{Growth} and $\mu=\cos\theta$, with $\theta$ the angle relative to the line-of-sight.

\item On small scales, galaxies have a velocity dispersion whose projection on the line-of-sight in redshift space gives rise to the "finger-of-God'' effect (FoG). These distortions can be modeled with a simple Gaussian orientation-dependent and scale-dependent damping model, which takes into account the pairwise peculiar-velocity dispersion of the galaxies $\sigma_{p}$  \cite{FoG}:
	\begin{equation}\label{eq:Dumbing}
		\ln D(k,\mu;\sigma_{p}) = { - \frac{1}{2} \left( \frac{k\sigma_{p}\mu}{H_{0}} \right)^{2}} \; .
	\end{equation}
This damping factor represents well actual data down to scales where $k\sigma_{p} \sim H_{0}$ \cite{FoG}, which corresponds to $r \sim 15h^{-1}\mathrm{Mpc}$. 
\end{itemize}

Finally, accounting for both effects and integrating over all orientations relative to the line-of-sight, the redshift-space galaxy power spectrum reads:
	\begin{equation} \label{eq:LSS_RSD}
		P^{(s)}_{gg}(k;b,\sigma_{p}) = b^{2} \int^{1}_{0} (1+\beta \mu^{2})^{2} D(k,\mu;\sigma_{p}) d\mu \ P^{(r)}_{\delta\delta}(k) \; ,
	\end{equation}
which can be analytically integrated over $\mu$, leading to:
\begin{eqnarray}
P^{(s)}_{gg}(k;b,\sigma_{p})=P^{(r)}_{\delta\delta}(k) \times \frac{b^2 H_0}{2k^5\sigma_p^5} \times  &\left[  -2\exp\left(-\frac{k^2 \sigma ^2}{2 H_0^2}\right) k \beta \sigma_p H_0 \left( k^2 (2+\beta) \sigma_p^2 + 3\beta H_0^2\right) \right. \nonumber \\
& \left. +\sqrt{2\pi} \mathrm{Erf}\left( \frac{k\sigma_p}{\sqrt{2}H_0}\right) \left(k^4 \sigma_p^4 + 2k^2\beta\sigma_p^2 H_0^2 + 3\beta^2 H_0^4\right)
\right] \; . \label{eq:model}
\end{eqnarray}	
Applying a Fast Fourier Transform to equation (\ref{eq:model}) results in the two-point correlation function with two parameters $b$ and $\sigma_p$:
	\begin{equation}\label{eq:fit-model}
		\xi^{(s)}(r;b,\sigma_{p}) = FFT\left[ P_{gg}^{(s)}(k;b,\sigma_{p})\right] \ .
	\end{equation}
Then we use equation (\ref{eq:Counts}) and (\ref{eq:Fractal}) to compute the $\Lambda$CDM prediction for $\mathcal{N}(<r)$ and $\mathcal{D}_{2}(r)$.

\subsection{Correction for bias and RSD}	
	
The estimators presented in section \ref{subsect:Estimators} measure the clustering properties of the galaxy distribution, we need to convert them into estimators that describe the clustering properties of the total matter distribution. We fit $\xi(r)$ in the range $1 \ h^{-1} < r < 40 \ h^{-1}$ Mpc with our model of equation~\ref{eq:fit-model} to obtain $b$ and $\sigma_p$.
If we take into account only the bias and the Kaiser effect, the angle-averaged correlation function is just multiplied by a constant factor, a squared effective bias. Since this factor is independent of $r$, the same factor applies to $\mathcal N(<r)-1$, which is an integral over $\xi(r')$ (see equation \ref{eq:Counts}). 
Taking into account also finger-of-God effect, the multiplicative factor is no longer independent of $r$. 
To transform the estimated  $\mathcal N(<r)-1$ for galaxies into an estimation for matter, we multiply it by the ratio of our model for $\mathcal N(<r)-1$ for matter (i.e.~$b=1$ and $\sigma_p=0$) to our model for the best fit value of $b$ and $\sigma_p$, as
	\begin{equation}\label{eq:convMatter}
		\widehat{\mathcal{N}}(<r) =\frac{ \mathcal{N}^{\rm model}(<r;b=1,\sigma_p=0) -1 }{ \mathcal{N}^{\rm model}(<r;b,\sigma_{p}) -1 } \times\left[ \widehat{\mathcal{N}}_{gal}(<r) - 1 \right]+ 1 \; .
	\end{equation} 
Then the fractal correlation dimension $\widehat{\mathcal{D}}_{2}(r)$ is obtained from $\widehat{\mathcal{N}}(<r)$ using equation \ref{eq:Fractal}.

Section~\ref{sec:RSDsensitivity} shows that taking into account the finger-of-God effect only contributes in a small change in the measurement of the homogeneity scale by typically 1\%. So the error due to the imperfection in the modelling of this effect is negligible.

\section{Results}{\vspace{0.1pt}}

In this section, we present the results of our analysis. We determine the range in comoving distance for which we can measure $\mathcal N(<r)$ with our sample. We quantify the uncertainties in our measurements using mock catalogues. We measure the homogeneity scale for the galaxy distribution,we fit the redshift-space-distortion parameters, measure the homogeneity scale for matter distribution. Finally we make comparisons with the $\Lambda$CDM model and estimate the average of $\mathcal D_2$ on large scales.

\subsection{Analysis range}

\begin{figure}[h!]
	\centering
	\includegraphics[width=100mm]{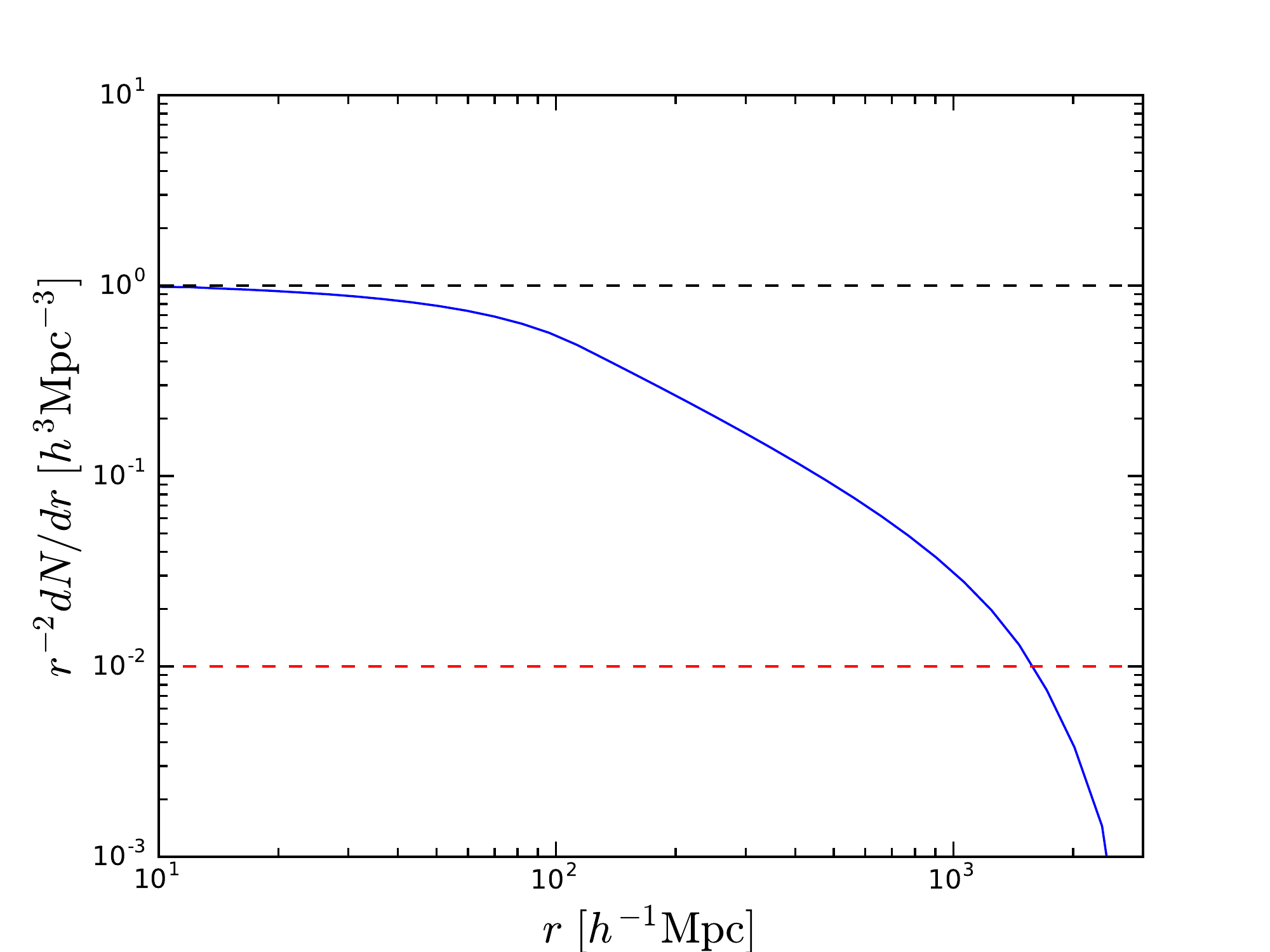}
												
	\caption{\label{fig:large_RR} Scaled number density of random pairs, $r^{-2}dN/dr$, versus the comoving radius of spheres, $r$, normalized to be unity at small $r$. }
	\end{figure}

Figure~\ref{fig:large_RR} shows the number density of random pairs, $dN/dr$, divided by $r^{2}$ and normalized such that at small scales it is unity. This scaled number density is constant at small scales and then decreases with $r$ due to the finite size of the survey. We analyse  data up to a maximum $r=1300~h^{-1}$ Mpc, where the scaled density goes down to $1\%$. 
	
\subsection{Covariance matrices}\label{subsec:CovMat}

We use a set of 1000 QPM mock catalogues to estimate the covariance matrices. We compute our observables for each of the catalogues and derive the bin-to-bin covariance matrices for each of the relevant observables, $\xi(r)$, $\mathcal{N}(<r)$ and $\mathcal{D}_{2}(r)$.

	\begin{figure}[h]
	\centering
	\includegraphics[width=.32\linewidth, keepaspectratio]{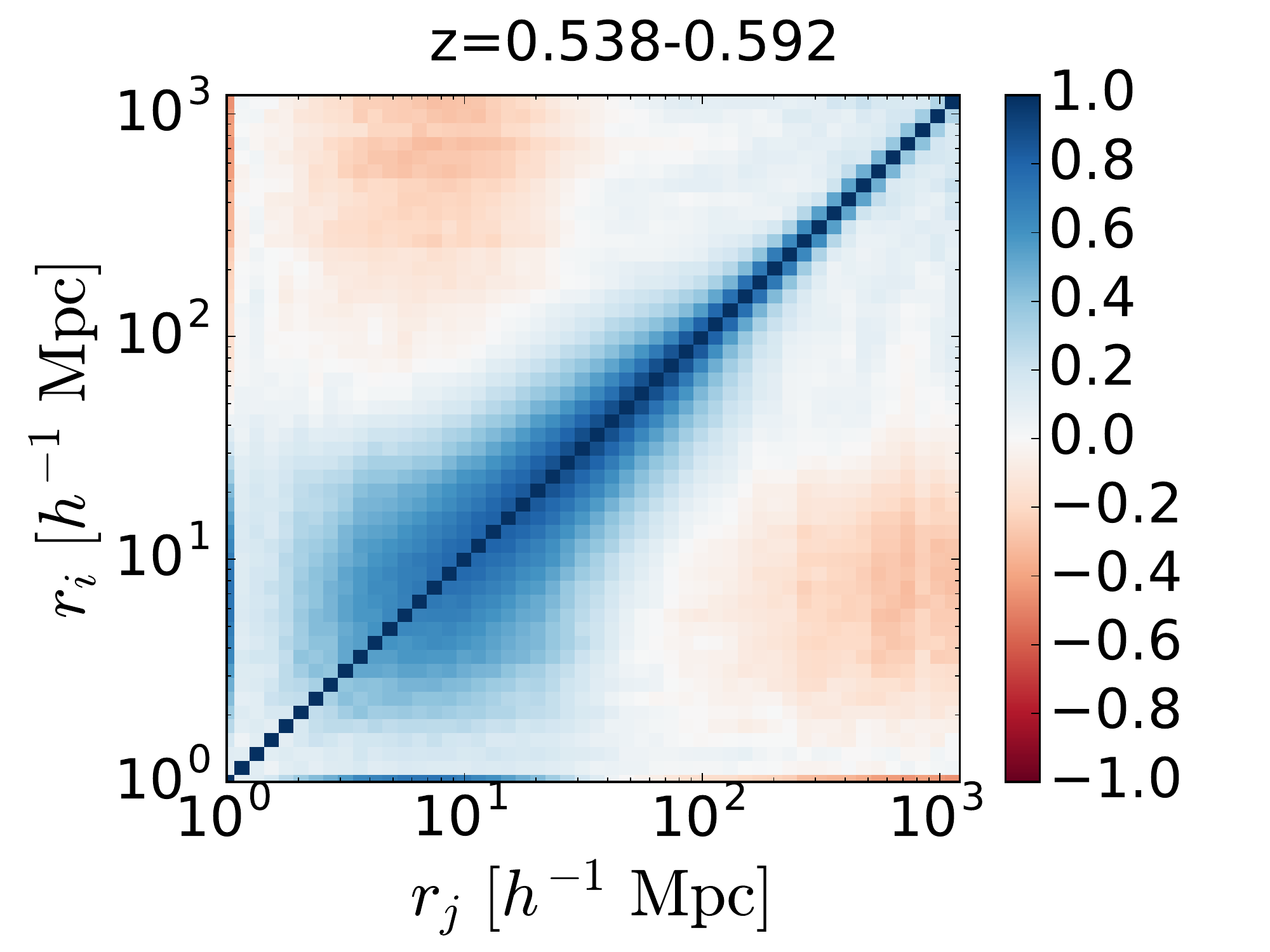}
	\includegraphics[width=.32\linewidth, keepaspectratio]{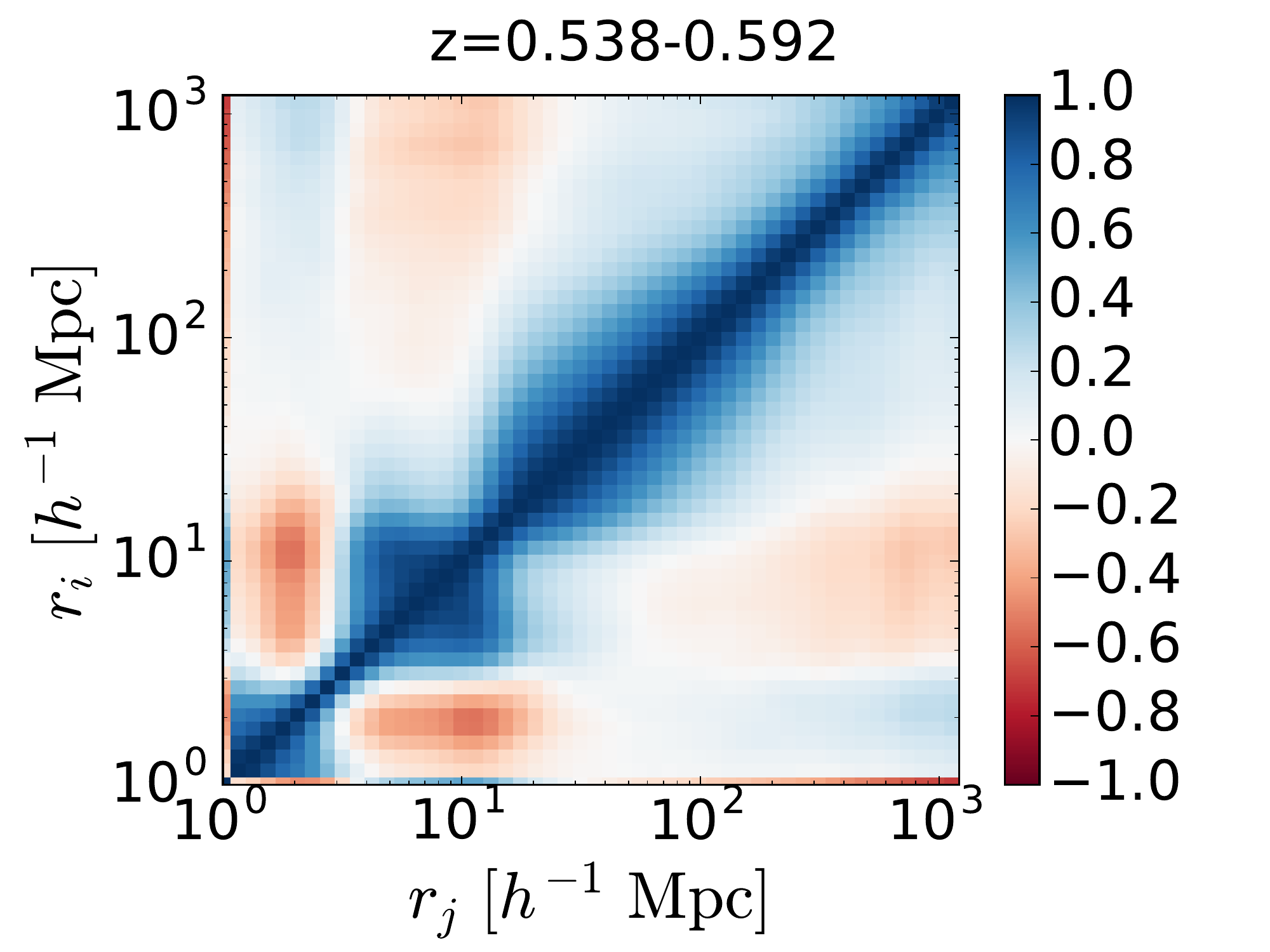}
	\includegraphics[width=.32\linewidth, keepaspectratio]{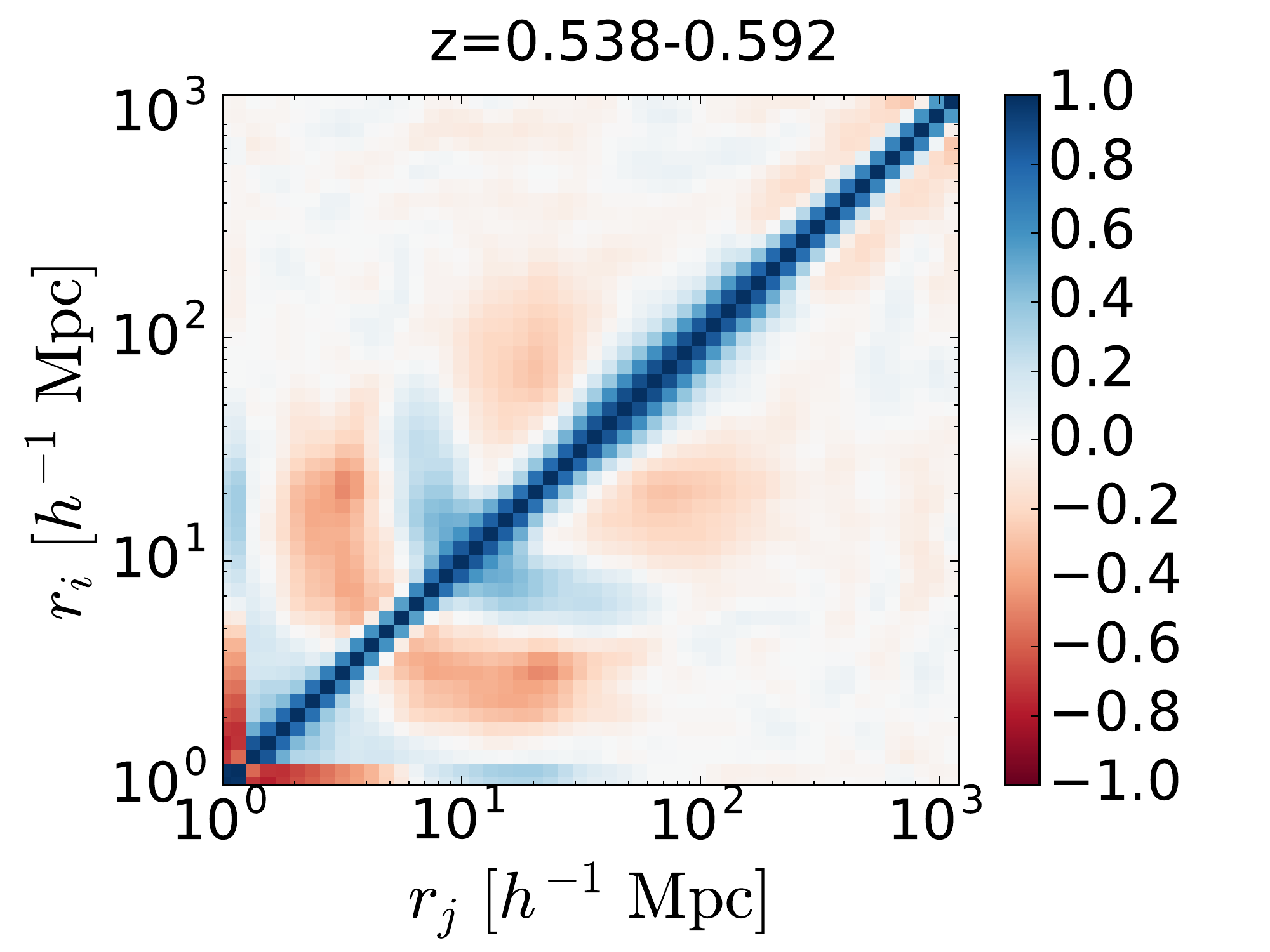}
	\caption{\label{fig:CorrMat} Correlation matrices for $\xi(r)$, $\mathcal{N}(<r)$ and $\mathcal{D}_{2}(r)$ for the total matter distribution in the redshift range $0.538 \le z \le 0.592$ .}
	\end{figure}

Fig.~\ref{fig:CorrMat} displays the resulting correlation matrices\footnote{We show the correlation matrices, $\rho_{ij} = \frac{C_{ij}}{\sqrt{C_{ii}*C_{jj}}}$, which have less dynamics than the covariance matrices.}
in the redshift bin $0.538 \le z \le 0.592$. The correlations matrices are similar in the other redshift bins.
$\mathcal{N}(r)$ is more correlated on large scales than $\xi(r)$ because it is an integral over $\xi(r)$.
On the other hand $\mathcal{D}_{2}(r)$, which is a derivative of $\mathcal{N}(r)$, is not very correlated.
Therefore, when studying the redshift evolution in section~\ref{subsec:RH}, we consider only the homogeneity scale obtained with $\mathcal{D}_{2}(r)$, and not with $\mathcal{N}(r)$.

\subsection{Homogeneity scale for the galaxy distribution}\label{subsec:RH-gal}

\begin{figure}[h!]
	\centering
	\includegraphics[width=.49\linewidth, keepaspectratio]{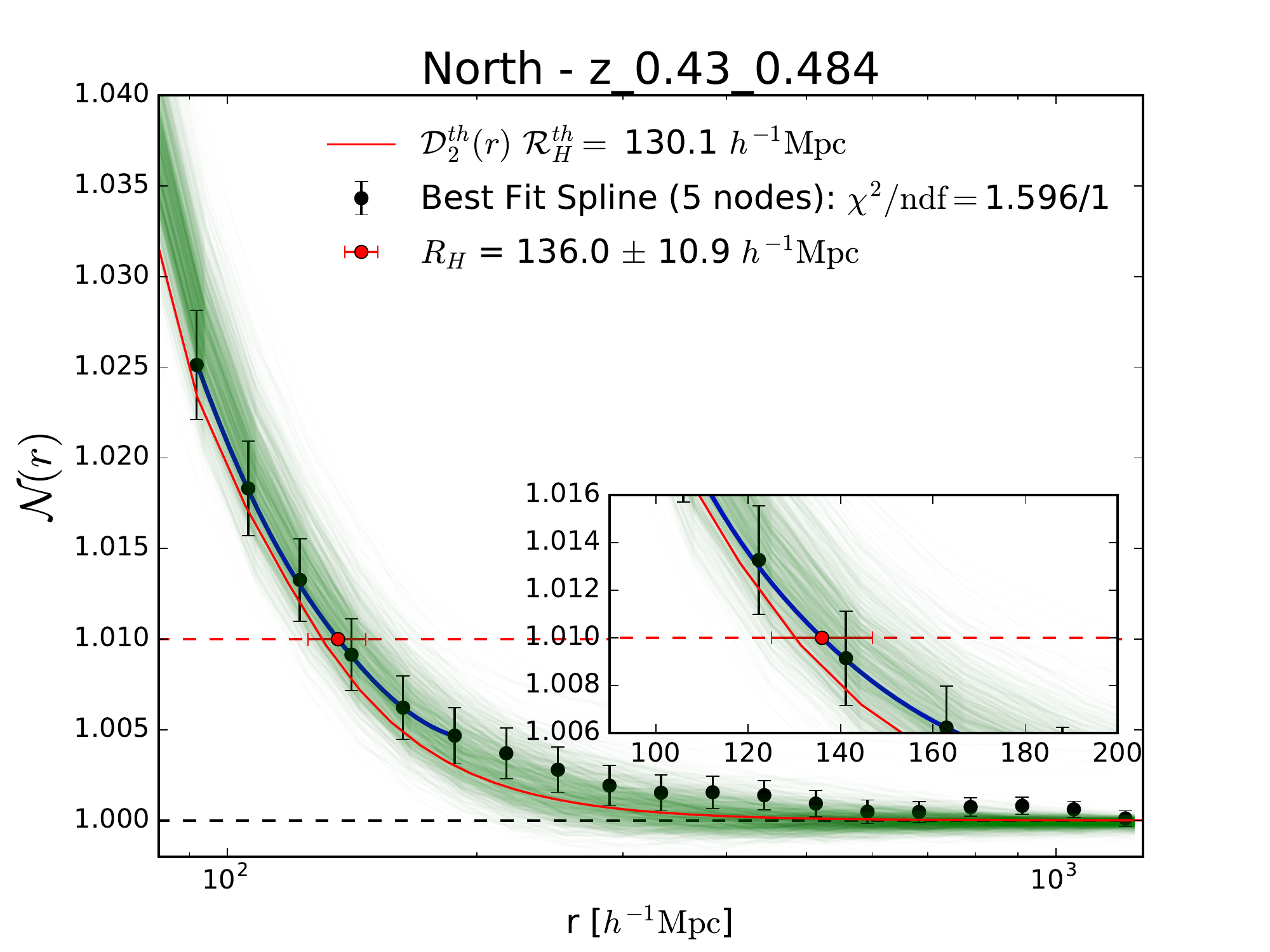}
	\includegraphics[width=.49\linewidth, keepaspectratio]{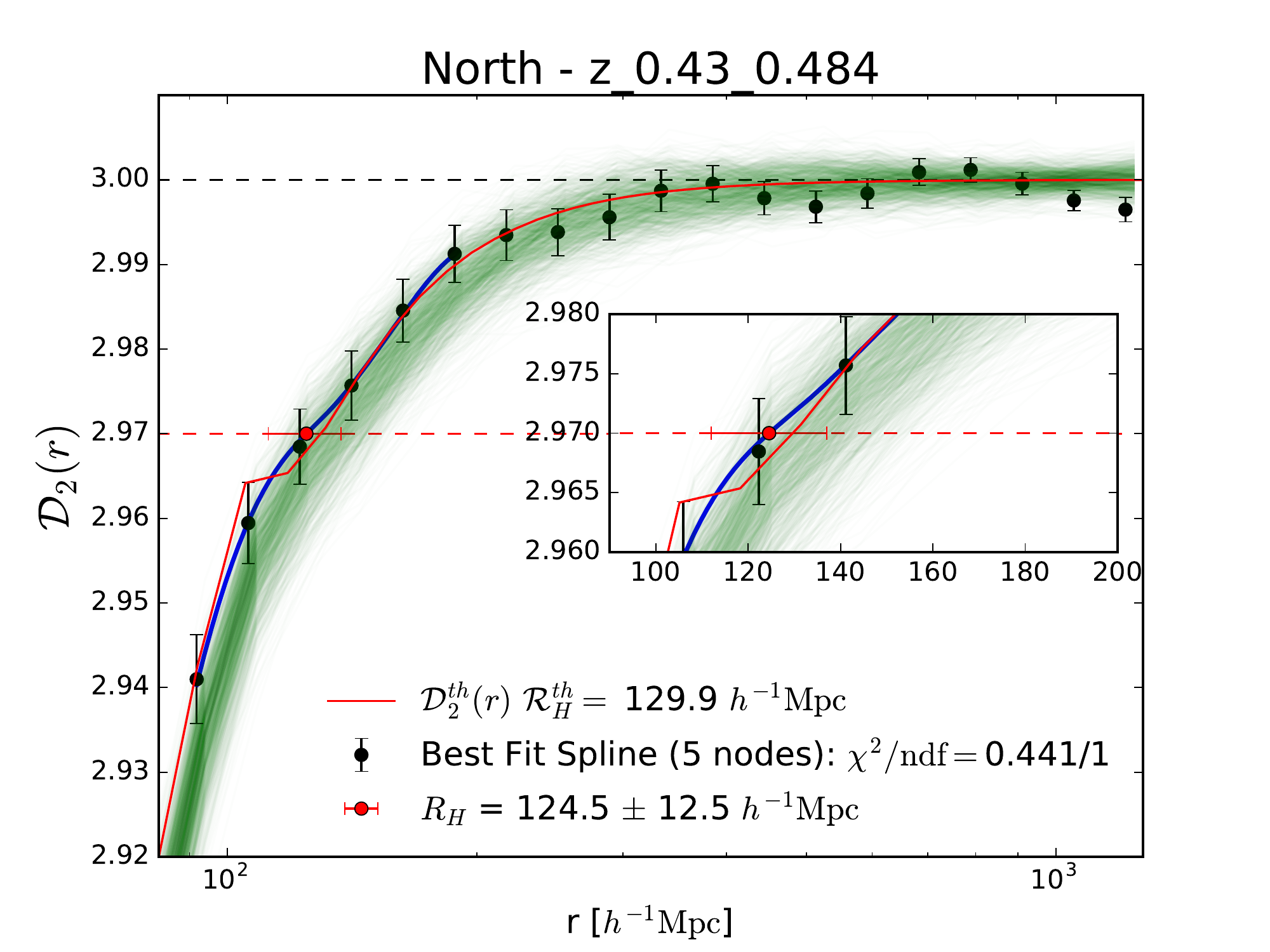}
	\caption{
\label{fig:RH-measure-GAL} Scaled counts-in-spheres, $\mathcal{N}(<r)$, (left) and fractal correlation dimension, $\mathcal{D}¯_{2}(r)$, (right). The black points with error bars are the result for the galaxy distribution in the NGC in the $0.430 \le z \le 0.484$ redshift bin. 
The blues lines are the best spline fit of the data.
The green lines are the results for the $1000$ QPM mock catalogues. 
The red continuous line is the $\Lambda$CDM prediction for $b=1.95$ and $\sigma_p=252\  $  km/s.
}
\end{figure}

We first compute  $\mathcal{N}^{gal}(<r)$  and $\mathcal{D}^{gal}_{2}(r)$ for the CMASS galaxy distribution using equation \ref{eq:Counts} and \ref{eq:Fractal}, without correcting for bias and redshift space distortions. This provides a measurement that does not rely on $\Lambda$CDM model to determine the bias. 
Results are shown in figure \ref{fig:RH-measure-GAL} for the $0.430 \le z \le 0.484$ redshift interval in the NGC.
As described in section~\ref{subsec:CovMat}, the error are obtained from the $1000$ QPM mock catalogues.

The horizontal black-dashed lines in figure \ref{fig:RH-measure-GAL} indicate the value of the corresponding observable for an homogeneous distribution. The two observables reach homogeneity on large scales. The intersection of the data with the red-dashed lines  (at $\mathcal N=1.01$ and $\mathcal D_2=2.97$) defines the homogeneity scale. In order to determine it, we perform a spline fit  over $6$ data point around the intersection. The number of nodes of the spline fit is set to $5$ in order to get an average $\chi^2$ equal to the number of degrees of freedom for the $1000$ QPM mocks. The error on $R_{H}$ was obtained from the errors in the spline factors by error propagation. The results, presented in Table~\ref{tab:RH-z} for the case of $\mathcal D_2$,  are consistent with our $\Lambda$CDM model predictions. We finally stress the fact that the behaviour of the data is consistent with that of the $1000$ QPM mock catalogues. 

\subsection{Determining the bias and the velocity dispersion}\label{subsec:RSD}
	
\begin{figure}[ht!]
	\setlength{\belowcaptionskip}{10pt}
	\centering
	\includegraphics[width=100mm]{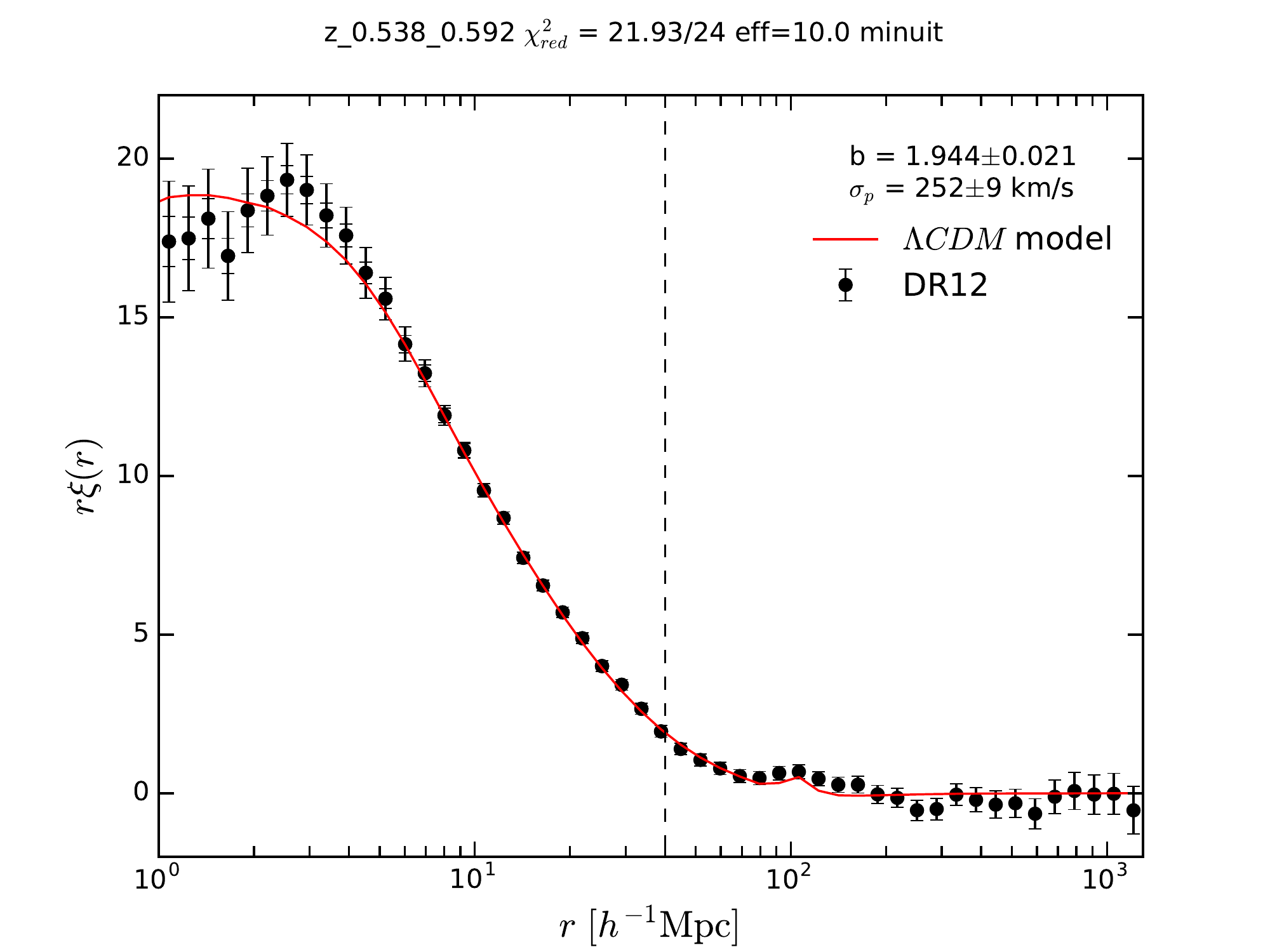}
\caption{\label{fig:bias} 
The correlation function of the CMASS galaxies in the $[0.538-0.592]$ redshift range. Data points have both their original error bars and the error bars enlarged to take into account the uncertainty of the RSD model on small scales 
(see section~\ref{subsec:RSD} and appendix~\ref{APP:RSD-robust}). 
The red line is the result of the fit performed over the range  $[1,40]\ h^{-1}$ Mpc (up to the vertical black-dashed line).
} 
\end{figure}

As illustrated in figure \ref{fig:bias}, we fit our model (Eq.~\ref{eq:fit-model}) 
to the measured correlation function for the CMASS galaxy sample, in order to determine the galaxy bias and velocity dispersion. As explained in Appendix \ref{APP:RSD-robust}, we account for theoretical uncertainties in the RSD modelling by boosting the covariance matrix on the relevant small scales. We thus achieve a good $\chi^2/$n.d.f. in the range $[1,40]\ h^{-1}$ Mpc. 
Appendix \ref{APP:RSD-robust} shows that this change in the covariance matrix at small scales has a negligible effect on the measurement of the homogeneity scale.	

Results are given in table \ref{tab:bias}. The mean precision  is 2.6\% (NGC) and 4.1\% (SGC) for the bias and 12\% (NGC) and 23\% (SGC) for the velocity dispersion. 
The bias is in agreement with values obtained by several authors \cite{FoG,bias2010,bias2012,ho2012clustering,bias_stochastic,gil2015power_bias} 
and the velocity dispersion is consistent with the typical CMASS-galaxy-sample velocity-dispersion, $\sigma_{p} \approx 240\pm50$ km/s~\cite{Sigma}.

\begin{table}[h!]
	\setlength{\abovecaptionskip}{-8pt}
		\begin{center} 
		\begin{tabular}{c|ccc|ccc} 
		\hline
		 & \multicolumn{3}{c}{NGC} & \multicolumn{3}{c}{SGC}\\
 		\hline
	 	$z$                   & $b$ & $\sigma_{p}\ $ [km/s] & $\chi_{red}^{2}$ & $b$ & $\sigma_{p}\ $ [km/s]  & $\chi_{\rm red}^{2}$ \\ 
	 	\hline 
		0.430-0.484 & $1.879 \pm 0.023$ & $243.0 \pm\ 9.7$   & 0.44 &\ \ $1.872 \pm 0.035$ & $238.7 \pm 15.2$   & 0.85 \\ 
	 	0.484-0.538 & $1.846 \pm 0.021$ & $247.1 \pm\ 7.4$   & 0.67 &\ \ $1.840 \pm 0.032$ & $234.6 \pm 12.1$   & 0.66 \\ 
	 	0.538-0.592 & $1.944 \pm 0.021$ & $252.0 \pm\ 8.9$   & 0.91 &\ \ $1.943 \pm 0.032$ & $236.2 \pm 13.7$   & 0.59 \\ 
		0.592-0.646 & $1.995 \pm 0.022$ & $234.2 \pm 10.9$   & 1.47 &\ \ $2.001 \pm 0.035$ & $237.9 \pm 19.6$   & 0.79 \\ 
		0.646-0.700 & $2.153 \pm 0.028$ & $235.3 \pm 21.7$   & 0.79 &\ \ $2.081 \pm 0.045$ & $210.3 \pm 40.1$   & 0.50 \\ 
	 	\hline
		\end{tabular}
		\end{center}
\caption{\label{tab:bias} Fitted values of bias, $b$, and velocity dispersion, $\sigma_p$,  in the different redshift bins, together with the corresponding reduced $\chi^{2}$ for $24$ degrees of freedom.}
\end{table}


\subsection{Homogeneity scale for the matter distribution}	\label{subsec:RH}

\begin{figure}[h!]
	\centering
	\includegraphics[width=.48\linewidth, keepaspectratio]{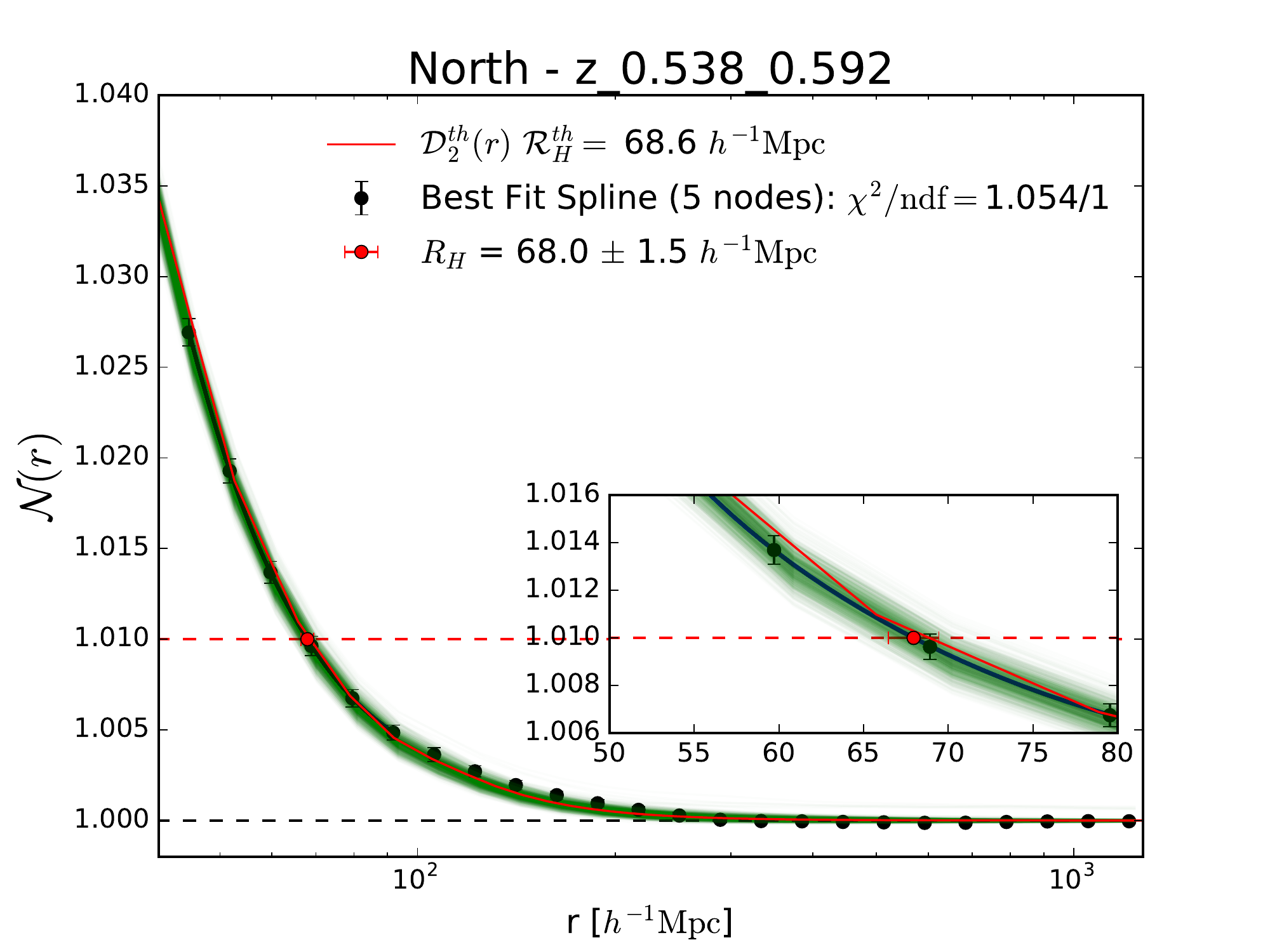}
	\includegraphics[width=.48\linewidth, keepaspectratio]{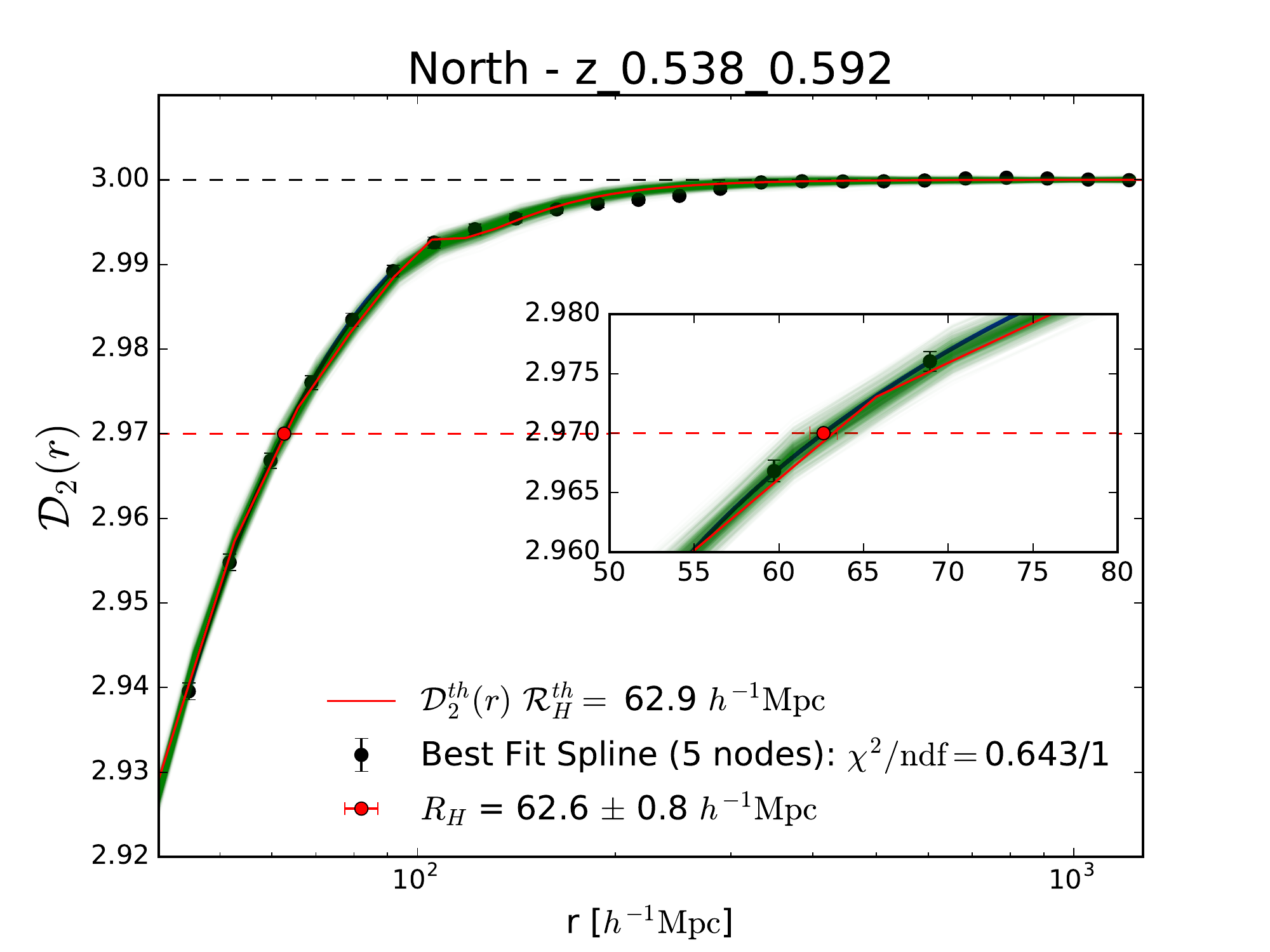}
\caption{\label{fig:RH-measure} 
Same as figure~\ref{fig:RH-measure-GAL} for matter distribution in the redshift interval $0.538 \le z \le 0.592$.}
\end{figure}

We use equation (\ref{eq:convMatter}) with parameter $b$ and $\sigma_p$ obtained in section \ref{subsec:RSD} to transform $\mathcal{N}^{gal}(<r)$ into $\mathcal{N}(<r)$ for matter. We then use Eq \ref{eq:Fractal}   
to get $\mathcal{D}_{2}(r)$ for matter distribution. Results are shown in figure \ref{fig:RH-measure} for the redshift interval $0.538 \le z \le 0.592$.
The two observables indicate homogeneity on large scales in this redshift interval, and in the four other intervals as well.
As in section~\ref{subsec:RH-gal} we fit the data points to determine the homogeneity scales. 
We stress that the fit range, $40<r<100\ h^{-1}$ Mpc, does not overlap with the fit range for determining the bias, $1<r<40\ h^{-1}$ Mpc.

The results, presented in Table~\ref{tab:RH-z} for the case of $\mathcal D_2$, are consistent with our $\Lambda$CDM model predictions. In the redshift interval $0.538\le z < 0.592$ we get a precision of 1.6\%, 
which is a factor $5$ better than~\citet{WiggleZ} in spite of  their wider redshift range, $0.5 < z< 0.7$.  
The more recent analysis by \citet{sarkar2016many} does not give $R_H$ for matter distribution.

\begin{figure}[h!]
	\centering
	\includegraphics[width=100mm]{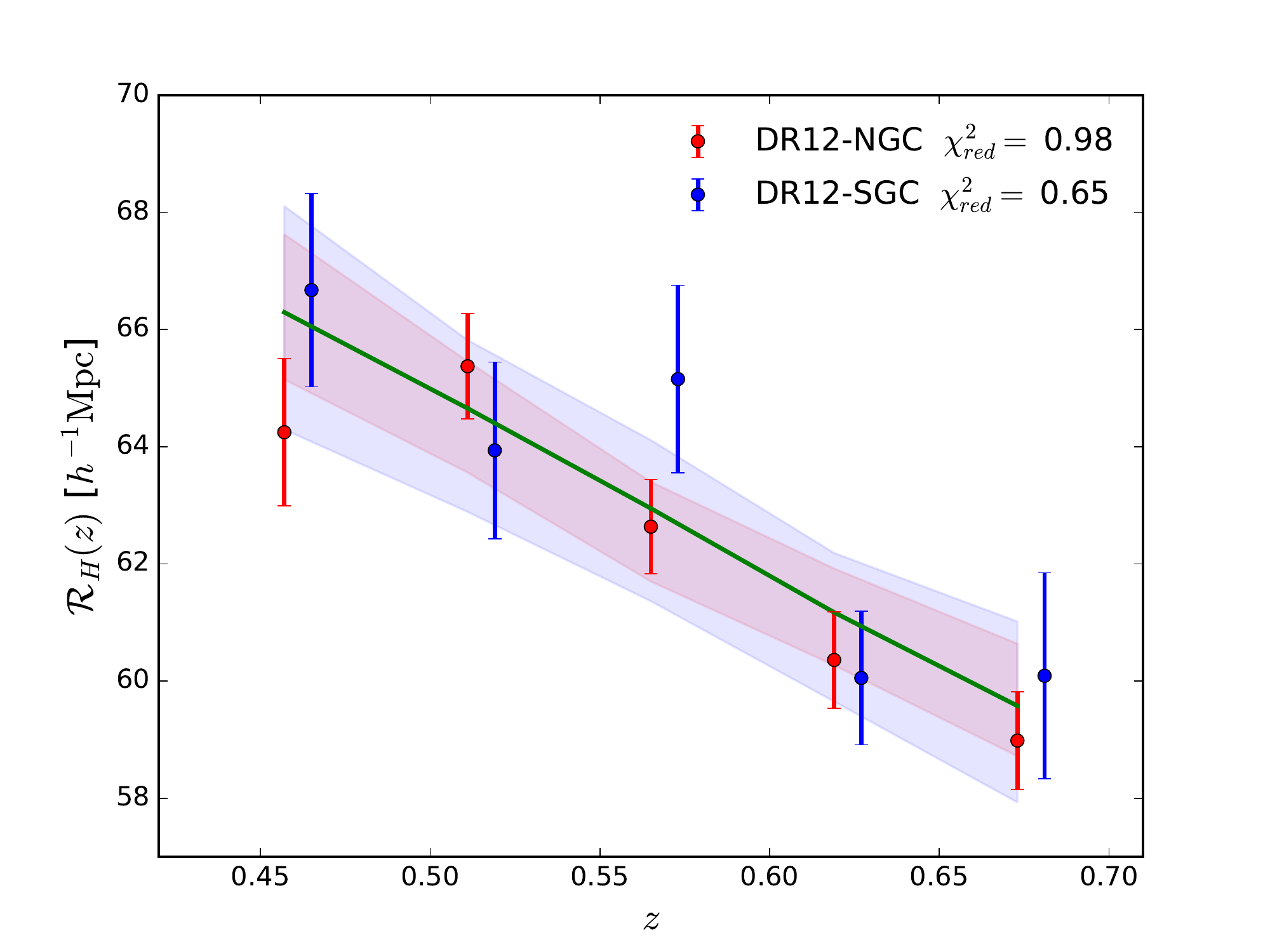}		
	\caption{\label{fig:RH-z}The homogeneity scale $R^{\mathcal{D}_2=2.97}_{H}(z)$ measured in the NGC (red) and in the SGC (blue)  as a function of redshift. The green line is the $\Lambda CDM$ model prediction.  The shaded areas indicate the $1\sigma$ range for the $1000$ QPM mock catalogues. }
\end{figure}
	 
Figure \ref{fig:RH-z} shows that the measured homogeneity scale is compatible with $\Lambda$CDM, with reduced $\chi^2$ smaller than unity both in the NGC and the SGC. The scale is increasing with time, from $z=0.7$ to $z=0.43$, as expected when clustering is increasing with time. 
The reduced $\chi^2$ remains smaller than unity when adding the result from~\citet{Laurent}, $\mathcal{R}_{H}(z=2.4)=26.2\pm0.9\ h^{-1}$Mpc.

\begin{table}[h!]
		\begin{center} 
		\begin{tabular}{c|cc|cc} 
		\hline
		 & \multicolumn{2}{c}{{$\mathcal{R}_{H}$[$h^{-1}$Mpc] Matter}} & \multicolumn{2}{c}{{$\mathcal{R}_{H}$[$h^{-1}$Mpc] Galaxy}}\\
 		\hline
	 	      $z$               &  NGC & SGC & NGC & SGC \\
		\hline 

		$0.430-0.484$ & $64.2 \pm 1.3  $ & $66.7 \pm 1.6 $	& $124.5 \pm 12.5  $ & $121.1 \pm\ 9.8 $	\\
		$0.484-0.538$ & $65.4 \pm 0.9  $ & $63.9 \pm 1.5 $	& $111.9 \pm\ 4.9  $ & $119.8 \pm\ 8.8 $	\\
		$0.538-0.592$ & $62.6 \pm 0.8  $ & $65.2 \pm 1.6 $	& $116.4 \pm\ 7.8  $ & $110.5 \pm\ 5.1 $	\\
		$0.592-0.646$ & $60.4 \pm 0.8  $ & $60.1 \pm 1.1 $	& $108.8 \pm\ 3.9  $ & $120.1 \pm 11.7 $	\\
		$0.646-0.700$ & $59.0 \pm 0.8  $ & $60.1 \pm 1.8 $  & $125.8 \pm\ 7.3  $ & $147.4 \pm\ 8.4 $ \\
	 	\hline
		\end{tabular}
		\end{center}
\caption{\label{tab:RH-z} Homogeneity scale, $R^{\mathcal{D}_2=2.97}_{H}(z)$, for the galaxy and matter distributions in the north and south galactic caps.}
\end{table}


\subsection{Consistency with $\Lambda$CDM} 

In order to characterize more precisely the agreement with the $\Lambda$CDM model, we fit $\mathcal{D}_2(r)$ from the data and the $1000$ mock catalogues in the range $40\ h^{-1} < r <1300\ h^{-1}$Mpc with 
	\begin{equation}\label{eq:modelA}
		 \mathcal{D}_2^{\Lambda\rm CDM} = \mathcal{D}_2(a r;\textbf{p}_{cosmo}) \; .
	\end{equation}
We fix the cosmological parameters $\textbf{p}_{cosmo}$ and leave only free the $a$.

\begin{figure}[h!]
	\centering
	\includegraphics[width=100mm]{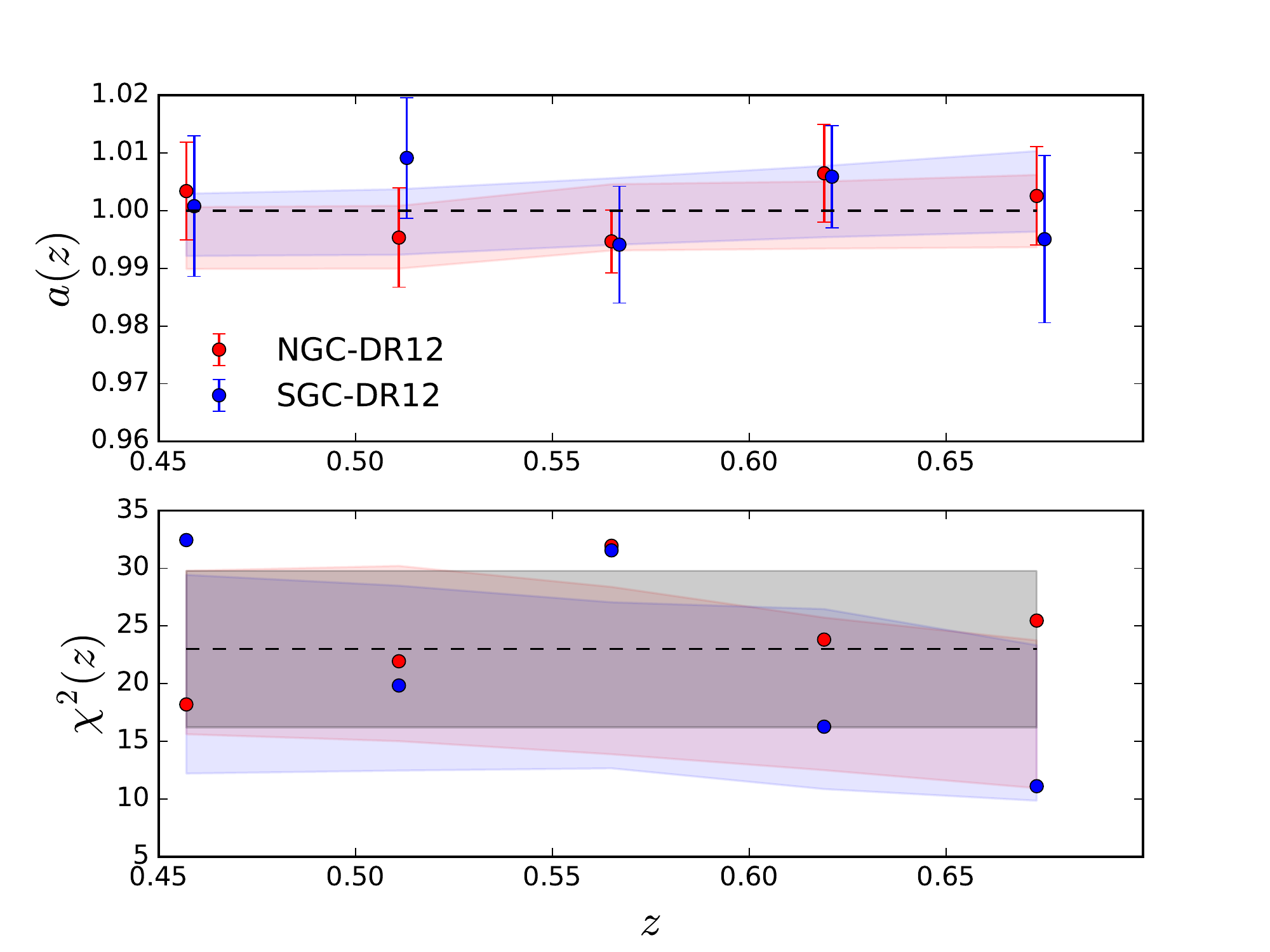}
	\caption{\label{fig:chi2-theory} 
Top: value of the parameter $\alpha$ resulting from the fit of the NGC (red) and SGC (blue) $\mathcal{D}_2(r)$ data with the model of equation  \ref{eq:modelA}, in different redshift bins. The shaded areas correspond to the $1\sigma$ region for the $1000$ QPM mock catalogues.
Bottom: The corresponding $\chi^2$.
}
\end{figure}

Figure \ref{fig:chi2-theory} shows the results of the fits in different redshift bins for both NGC and SGC.
The values of  $a$, all consistent with $1$, demonstrate a good agreement with $\Lambda$CDM model. 
Furthermore, the bottom plot of the figure shows that, for the data, the $\chi^2$ are consistent with the number of degrees of freedom (23), as illustrated by the  grey shaded area that indicates the expected $1\sigma$ extension, $\chi^2= 23 \pm \sqrt{2\times23}$. 
The mock catalogues, however, are a bit at variance with this grey area at large $z$. 
		
\subsection{Constraints on fractal correlation dimension}\label{subsect:ConD2}

\begin{figure}[h!]
	\centering
	\includegraphics[width=100mm]{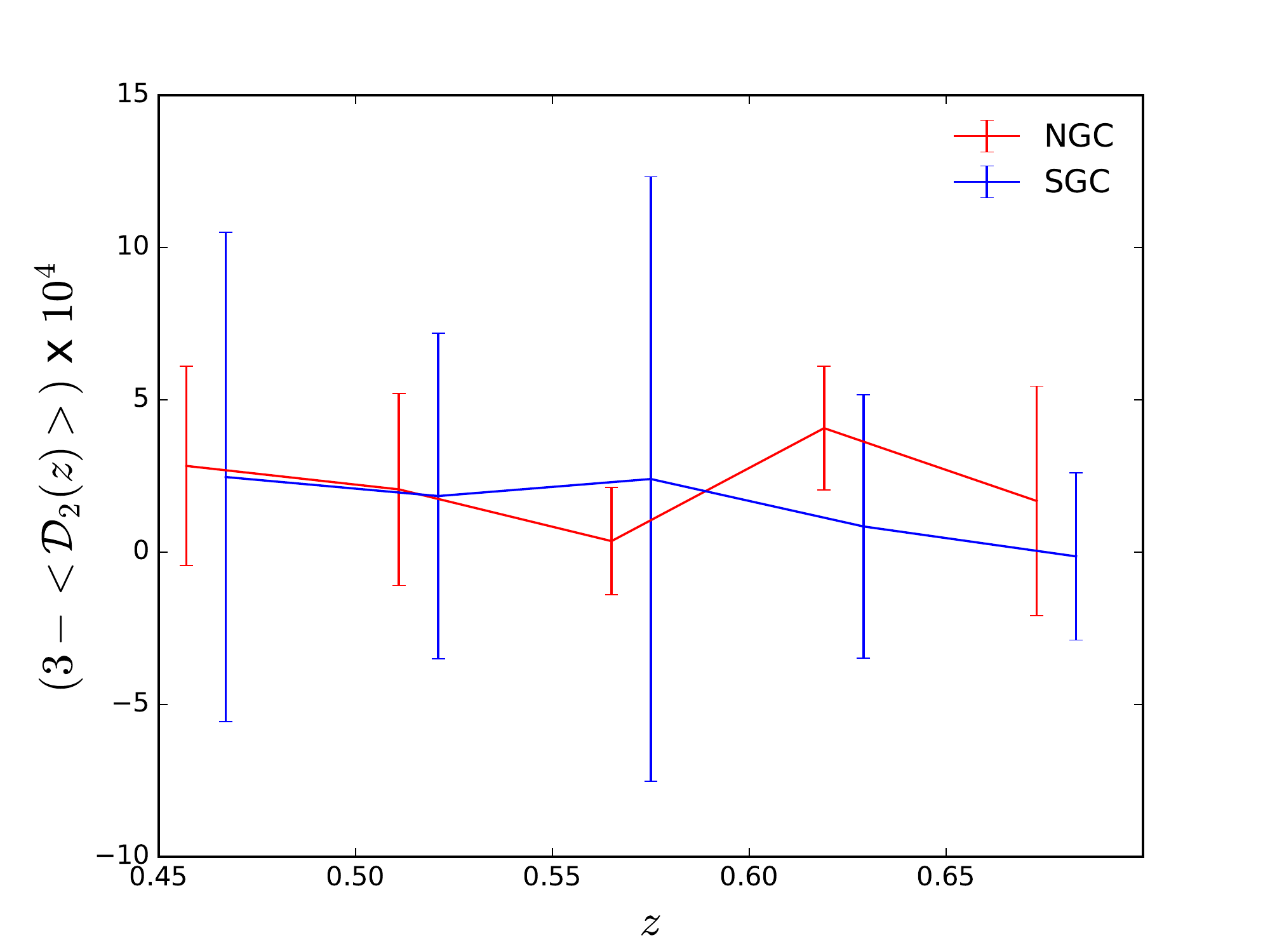}
	\caption{\label{fig:D2zero} Fractal correlation dimension for the matter distribution, $3-\langle \mathcal{D}_2 \rangle$, averaged over $300\ h^{-1} < r < 1300\ h^{-1}$ Mpc, in the different redshift bins.}	
\end{figure}

\begin{table}[h!]
		\begin{center} 
		\begin{tabular}{*5c} 
		\hline
			 & \multicolumn{2}{c}{{Matter}} & \multicolumn{2}{c}{{Galaxy}}\\
 		\hline
	 	      $z$               &  NGC $(\times 10^{-4})$ & SGC $(\times 10^{-4})$ & NGC $(\times 10^{-3})$ & SGC $(\times 10^{-3})$ \\
		\hline 

		$0.430-0.484$ & $2.8 \pm 3.3  $ & $\ \ 2.5 \pm 8.1 $	& $1.3 \pm 1.5  $ & $1.1 \pm 3.7 $	\\
		$0.484-0.538$ & $2.0 \pm 3.1  $ & $\ \ 1.9 \pm 5.4 $	& $0.9 \pm 1.4  $ & $0.8 \pm 2.4 $	\\
		$0.538-0.592$ & $3.6 \pm 1.7  $ & $\ \ 2.4 \pm 9.9 $	& $0.2 \pm 0.9  $ & $1.2 \pm 4.9 $	\\
		$0.592-0.646$ & $4.1 \pm 2.0  $ & $\ \ 0.9 \pm 4.5 $	& $2.1 \pm 1.1  $ & $0.5 \pm 2.3 $	\\
		$0.646-0.700$ & $1.7 \pm 3.8  $ & $-0.2 \pm 2.8 $  & $1.0 \pm 2.3  $ & $0.1 \pm 1.6 $ \\
	 	\hline
		\end{tabular}		
	\end{center}
\caption{\label{tab:MeanD2} Fractal correlation dimension, $3-\langle\mathcal{D}_{2}\rangle_{r}(z)$, averaged over  $300\ h^{-1} < r < 1300\ h^{-1}$ Mpc, with 1$\sigma$ errors, in the NGC and SGC.}
\end{table}

In order to assess the level of homogeneity of the CMASS DR12 galaxy sample we compute the average of $\mathcal D_2(r)$ over the range 
 $300\ h^{-1} < r < 1300\ h^{-1}$Mpc, accounting for the covariance matrix.
 The results are presented in table \ref{tab:MeanD2} for the different redshift bins in the NGC and SGC.
All results are compatible and we average them to get $3-\langle\mathcal{D}_{2}\rangle_{r,z} = (0.9 \pm 1.2)\times 10^{-3} $ ($1\sigma$).
 
Strictly speaking, it is not possible to transform the results for the galaxy distribution into results for matter distribution without using $\Lambda$CDM prediction to compute the galaxy bias. However,  \citet{Laurent} show that, with a set of reasonable assumptions independent of $\Lambda$CDM, it is possible to obtain a lower limit for the tracer bias. In our case this results in $b > \sqrt{1.6}$ and $3 - \langle \mathcal{D}_2 \rangle <  7.12 \pm 9.48 \times 10^{-4}$ for matter distribution.
 
Alternatively we can assume $\Lambda$CDM, use it to get the galaxy bias and compute $\mathcal{D}_2(r)$. We average $\mathcal{D}_2(r)$ over  $300\ h^{-1} < r < 1300\ h^{-1}$ Mpc and give the results in table \ref{tab:MeanD2} and in figure \ref{fig:D2zero}. Since the galaxy bias is significantly larger than one, the constraints are tighter for matter distribution than for galaxy distribution. Averaging over redshift and caps we get $3-\langle\mathcal{D}_{2}\rangle_{r,z} = (1.7 \pm 1.0)\times 10^{-4} $ ($1\sigma$). This is a strong consistency check of $\Lambda$CDM.

Using a similar estimator to \citet{Yadav1} on $1000$ QPM mock catalogues, we find a minimum  $160\ h^{-1}\mathrm{Mpc}$, maximum $1250\ h^{-1}\mathrm{Mpc}$, a spread of $120\ h^{-1}\mathrm{Mpc}$ and an average of $320\ h^{-1}\mathrm{Mpc}$ corresponding to a precision of about $30\%$ which is less precise than the results, we present in this study. A KS-test \cite{KS_test} for this estimator suggest that data and mocks are drawn from the same distribution at $21\%$ and $83\%$ C.L. for NGC and SGC respectively.

\section{Analysis Robustness}

\subsection{Bias in the fractal correlation dimension}

We use homogeneous random catalogues to take into account the inhomogeneity of the survey,
so we may wonder if this could bias the resulting  $\mathcal N(<r)$ towards homogeneity.
To search for such a possible bias, we generate $500$ fractal realizations with a given value of the fractal correlation dimension,
we pass them through our pipeline analysis and study the resulting $\mathcal{D}_2$, 
as done by \citet{Laurent}.
	 
Following~\citet{castagnoli1991}, we create a cubic box of $L\approx4\ h^{-1}$ Gpc side, containing the whole survey at $z=0.538-0.592$. We divide this box in $M=n^3$ sub-boxes of size $L/n$, where $n=2$ and we give to each sub-box a survival probability $p$. We then repeat the procedure for each surviving sub-box.  An infinite number of iterations would give a fractal distribution with:
	\small{\begin{equation}
		D_{2} =  \frac{\log(pM)}{\log(n)}
	\end{equation} }
	
We perform $9$ iterations.  After the last iteration, we populate each sub-box with random points that follow a Poisson law of mean $\lambda<1$. Then we convert the cartesian coordinates to $z,RA,DEC$ with the same FLRW metric as used in our analysis. Finally, we apply cuts to simulate the selection function of our galaxy survey. The value of $\lambda$ is chosen so that after cuts the number of objects in the fractal distribution is approximately the same as in our survey. 
	
	\begin{figure}[h!]
	\centering
	\includegraphics[width=100mm]{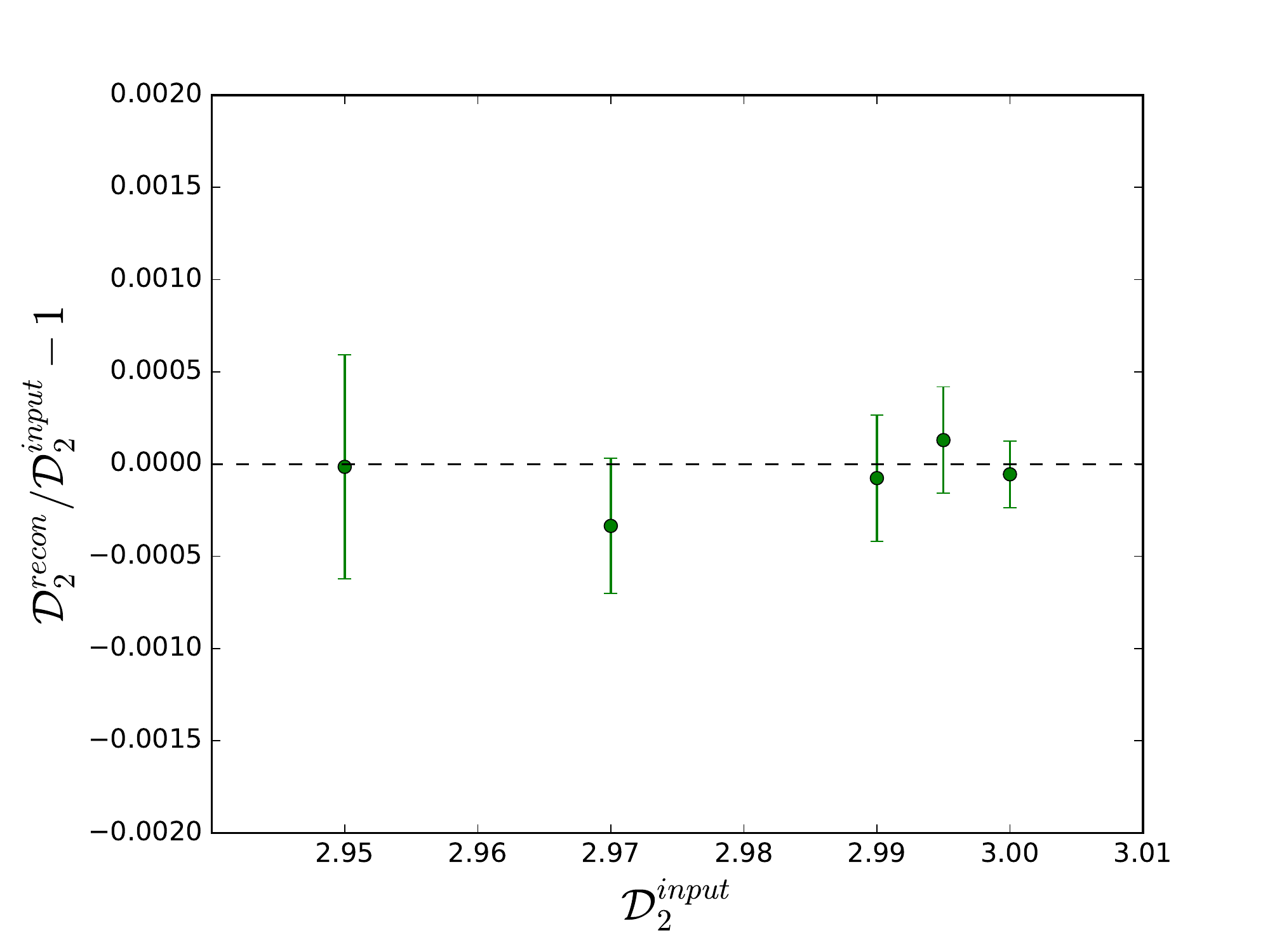}
	\caption{\label{fig:fractal} Reconstructed fractal correlation dimension, $\mathcal{D}_2$, in the redshift interval $z=0.538-0.592$, averaged over $r \in [15,1300]\ h^{-1}$ Mpc.}
	\end{figure}

At the last iteration the size of the sub-box is 15 $h^{-1}$ Mpc. So we average the reconstructed $\mathcal{D}_2$ over $r\in [15,1300]\ h^{-1}$ Mpc.
Figure \ref{fig:fractal} presents this average for different input $\mathcal{D}_2$. Fitting these points with a straight line, $y=\alpha x+\beta$, results in $\alpha=1.00\pm0.03$ and $\beta=(-0.4\pm7.9)\times 10^{-2}$ with $\chi_{\rm red}^2=0.9/3$. So, in contrast with~\citet{WiggleZ}, we do not observe any bias in the reconstructed fractal correlation dimension.

\newpage

\subsection{Sensitivity to RSD model}
\label{sec:RSDsensitivity}

We compare the homogeneity scale obtained with two different modelling of redshift space distortions. 
Namely, the model with a scale-independent bias and the Kaiser effect, see equation (\ref{eq:largeRDS}) :
	\begin{equation}\label{eq:RSD2}
		\xi(r;b) = FFT\left[ P_{gg}(k,\mu;b) \right] \; ,
	\end{equation}	
and our full model, which includes in addition a modeling of the finger-of-God effect (Eq.~\ref{eq:LSS_RSD}): 
	\begin{equation}\label{eq:RSD3}
		\xi(r;b,\sigma_p) = FFT \left[ P_{gg}(k,\mu;b,\sigma_p) \right]
	\end{equation}
	\begin{figure}[t]
	\centering
	\includegraphics[width=120mm]{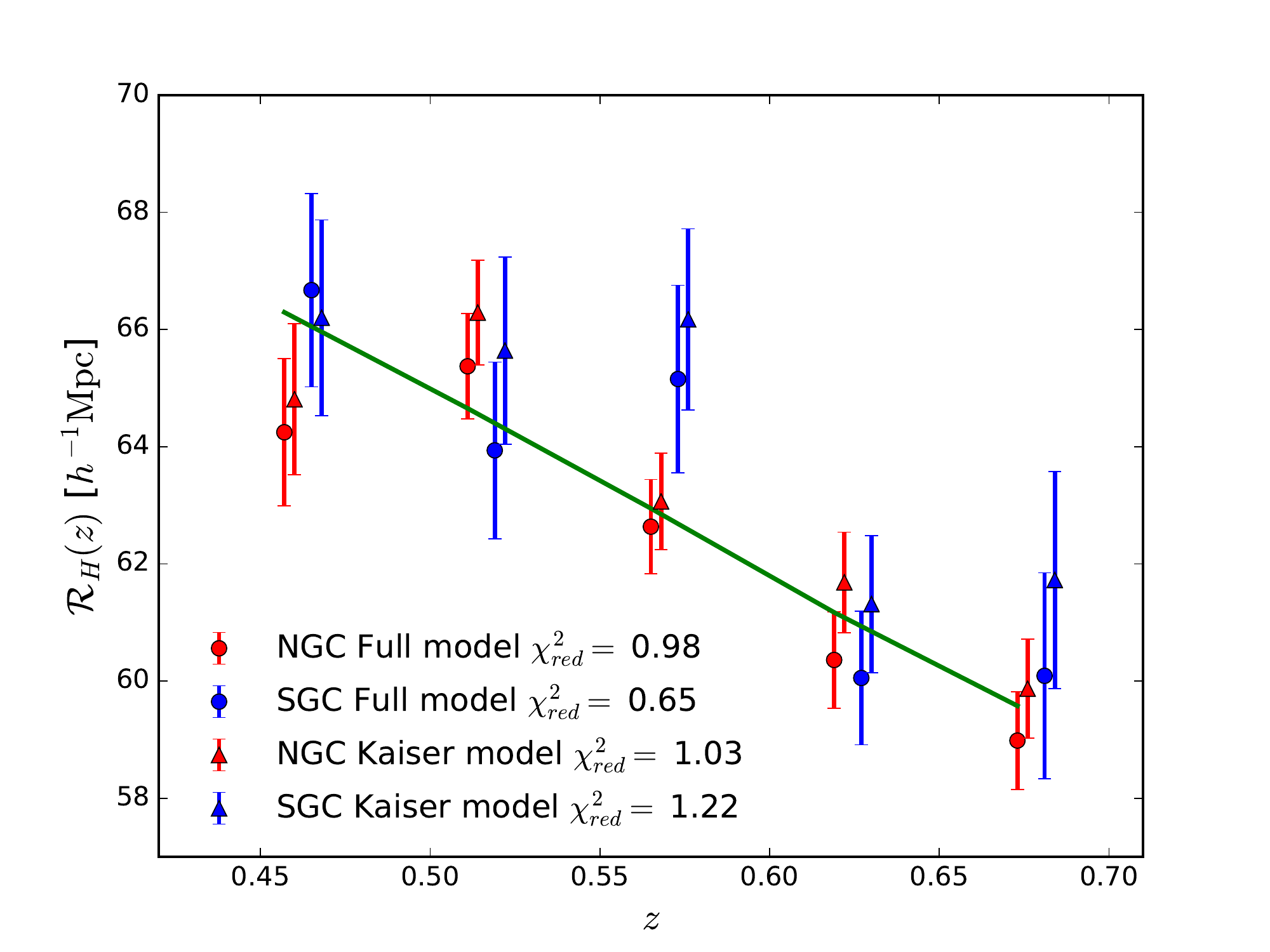}
	\caption{\label{fig:RSD_model_diff} Measured homogeneity scale in different redshift bins, in the NGC and the SGC, with Kaiser (triangle) and full RSD model (circles).}
	\end{figure}

Figure $\ref{fig:RSD_model_diff}$ presents the homogeneity scale for these two models in the five redshift bins for the north and south galactic caps. We note that using the full RSD model lowers  the $\chi^2$, confirming that this is a better model than the purely linear Kaiser model. The full model tends to lower $R_H$, by up to slightly more than 1$\sigma$, or about 1\%. This means that we should not use the purely linear Kaiser model. The remaining error due to the imperfection of our full RSD model is most probably a fraction of the difference between Kaiser and full models, and therefore negligible with respect to our statistical error.   
\section{Conclusions}

We use the data release 12  of BOSS CMASS galaxy sample to study the transition to cosmic homogeneity over a volume of 5.1 $h^{-3}$ Gpc$^3$. 
We do not consider the correlation function, $\xi(r)$, to study homogeneity because its definition requires an average density, which is only defined for a homogeneous sample. We rather use the counts-in-spheres, $N(<r)$, i.e.~the average number of objects around a given object, and its logarithmic derivative, the fractal correlation dimension, $D_2(r)$. For a homogeneous sample, $N(<r) \propto r^3$ and $D_2=3$. We define a characteristic homogeneity scale, $\mathcal{R}_H$, as the value for which $D_2$ reaches the homogeneous value within 1\%, i.e.~$D_2(\mathcal{R}_H)$=2.97. 

For the distribution of galaxies, we get $3 - \langle  D_2 \rangle = (0.6 \pm 1.3) \times 10^{-3}$ at 1 $\sigma$ over the range  $300\ h^{-1} \le r \le 1300\ h^{-1}$ Mpc, consistent with homogeneity and a transition to homogeneity at $\mathcal{R}_H = 114.2 \pm 5.8$  $h^{-1}$ Mpc. However, our analysis makes use of a random catalog to take into account the geometry and the completeness of the survey. The redshift distribution of this catalogue is taken from the data. We are therefore insensitive to a possible isotropic variation of the density with redshift, $\rho=\rho(z)$. In other words, we can only check for spatial isotropy, $\rho(z,\theta_1)=\rho(z,\theta_2)$. We stress that the same is true for all galaxy redshift surveys~\cite{Mustapha}.
On the other hand, this spatial isotropy can be obtained without using any fiducial model as discussed by \citet{Laurent}. So if we assume the Copernican principle our data imply homogeneity of the galaxy sample without any $\Lambda$CDM assumption.

Using an estimator similar to that of \citet{Yadav1}, we find agreement between the mocks and data, further confirming the transition to homogeneity in the matter and galaxy distributions, despite the lower precision of this estimator ($30\%$ compared to our estimator which is near $\%$-level). This estimator provides a qualitative estimate of the homogeneity scale in the range of $150\ h^{-1} \le r \le 400\ h^{-1}\  Mpc$. 

Alternatively, we can make a cross check of $\Lambda$CDM model. Thus, we fit  the CMASS galaxy correlation function in the range $1\ h^{-1} < r < 40\ h^{-1}\mathrm{Mpc}$ to obtain the galaxy bias relative to the $\Lambda$CDM prediction for the matter correlation function. We correct our measurement of $N(<r)$ for this bias in order to get the result for the matter distribution, that we finally compare to the $\Lambda$CDM prediction. 

For the matter distribution, we get $3-\langle \mathcal{D}_{2}({r>300h^{-1}\mathrm{Mpc}}) \rangle_{z} = (1.7\pm1.0) \times 10^{-4}  $ at $1\sigma$ and a transition to homogeneity at a characteristic scale $\mathcal{R}_H= 61.9 \pm 0.8 \ h^{-1}\mathrm{Mpc}$ at an average $z= 0.538-0.592$. This measurement of $\mathcal{R}_H$ is more precise than previous measurement by ~\citet{WiggleZ} by a factor $5$, while \citet{sarkar2016many}, only give a qualitative measurement of $\mathcal{R}_H$. 
We also investigate the redshift evolution of our observables. We find that the homogeneity scale is decreasing with time as expected if clustering is increasing with time. We find accordance with $\Lambda CDM$ model with a reduced $\chi^2 = 0.89(0.61)$ for the North(South) Galactic Cap for the $6$ redshift bins.

\section*{Aknowledgements}
	We would like to thank Jean-Philippe Uzan, Cyril Pitrou and Ruth Durrer for useful suggestions and discussions. PN would like to thank James Rich, although a coauthor, for fruitful discussions on the understanding of this work. PN would like also to thank Mikhail Stolpovskiy and Ranajoy Banerji for useful suggestion on the computational side of this project.
	
	Funding for SDSS-III has been provided by the Alfred P. Sloan Foundation, the Participating Institutions, the National Science Foundation, and the U.S. Department of Energy Office of Science.
The SDSS-III web site is http://www.sdss3.org/.
	
	SDSS-III is managed by the Astrophysical Research Consortium for the Participating Institutions of the SDSS-III Collaboration including the University of Arizona, the Brazilian Participation Group, Brookhaven National Laboratory, Carnegie Mellon University, University of Florida, the French Participation Group, the German Participation Group, Harvard University, the Instituto de Astrofisica de Canarias, the Michigan State/Notre Dame/JINA Participation Group, Johns Hopkins University, Lawrence Berkeley National Laboratory, Max Planck Institute for Astrophysics, Max Planck Institute for Extraterrestrial Physics, New Mexico State University, New York University, Ohio State University, Pennsylvania State University, University of Portsmouth, Princeton University, the Spanish Participation Group, University of Tokyo, University of Utah, Vanderbilt University, University of Virginia, University of Washington, and Yale University.
	
	MV is partially supported by Programa de Apoyo a Proyectos de Investigaci\'on e Innovaci\'on Tecnol\'ogica (PAPITT) No IA102516, Proyecto Conacyt Fronteras No 281 and Proyecto DGTIC SC16-1-S-120.
	
	GR acknowledges support from the National Research Foundation of Korea (NRF) through NRF-SGER 2014055950 funded by the Korean Ministry of Education, Science and Technology (MoEST), and from the faculty research fund of Sejong University in 2016
	
	This research used resources of the National Energy Research Scientific Computing Center, a DOE Office of Science User Facility supported by the Office of Science of the U.S. Department of Energy under Contract No. DE-AC02-05CH11231.
\appendix
\renewcommand{\thesection}{\Alph{section}}

\section{From $\xi(r)$ to $\mathcal{N}(<r)$}\label{APP:Nr_of_xi}

The probability of finding a galaxy within a volume $dV$ around another galaxy depends on the two-point correlation function $\xi(\vec{r})$~\cite{LSS-Peebles}:
	\begin{equation}
		dP = \bar{\rho} \left[ 1 + \xi(\vec{r}) \right] dV \; .
	\end{equation}
The {counts-in-spheres} of the distribution of galaxies is then related to the correlation function:
\begin{equation}
	N(<r) = \int dP = \bar{\rho} \int \left[ 1+\xi(\vec{r'}) \right] dV \ .
\end{equation}
Assuming $\xi(\vec{r})=\xi(r)$ we get:
\begin{equation}\label{count_in_spheres_app}
	N(<r)  =  4\pi\bar{\rho} \int^{r}_{o} \left[ 1+\xi(r') \right] r'^{2} dr' \ .
\end{equation}
For the random homogeneous distribution, $\displaystyle	N_{R}(<r) = \bar{\rho} \frac{4\pi}{3}r^{3}$,
so
\begin{equation}
	\mathcal{N}(<r) = \frac{N(<r)}{N_R(<r)} = \frac{3}{r^3}\int^{r}_{o} \left[ 1+\xi(r') \right] r'^{2} dr' = 1 + \frac{3}{r^3}\int^{r}_{o} \xi(r') r'^{2} dr' \ .
\end{equation}

\section{Choice of estimator for ${\mathcal{N}}(<r)$} \label{subsec:Choice}

\begin{figure}[h!]
	\centering
	\includegraphics[width=0.48\linewidth, keepaspectratio]{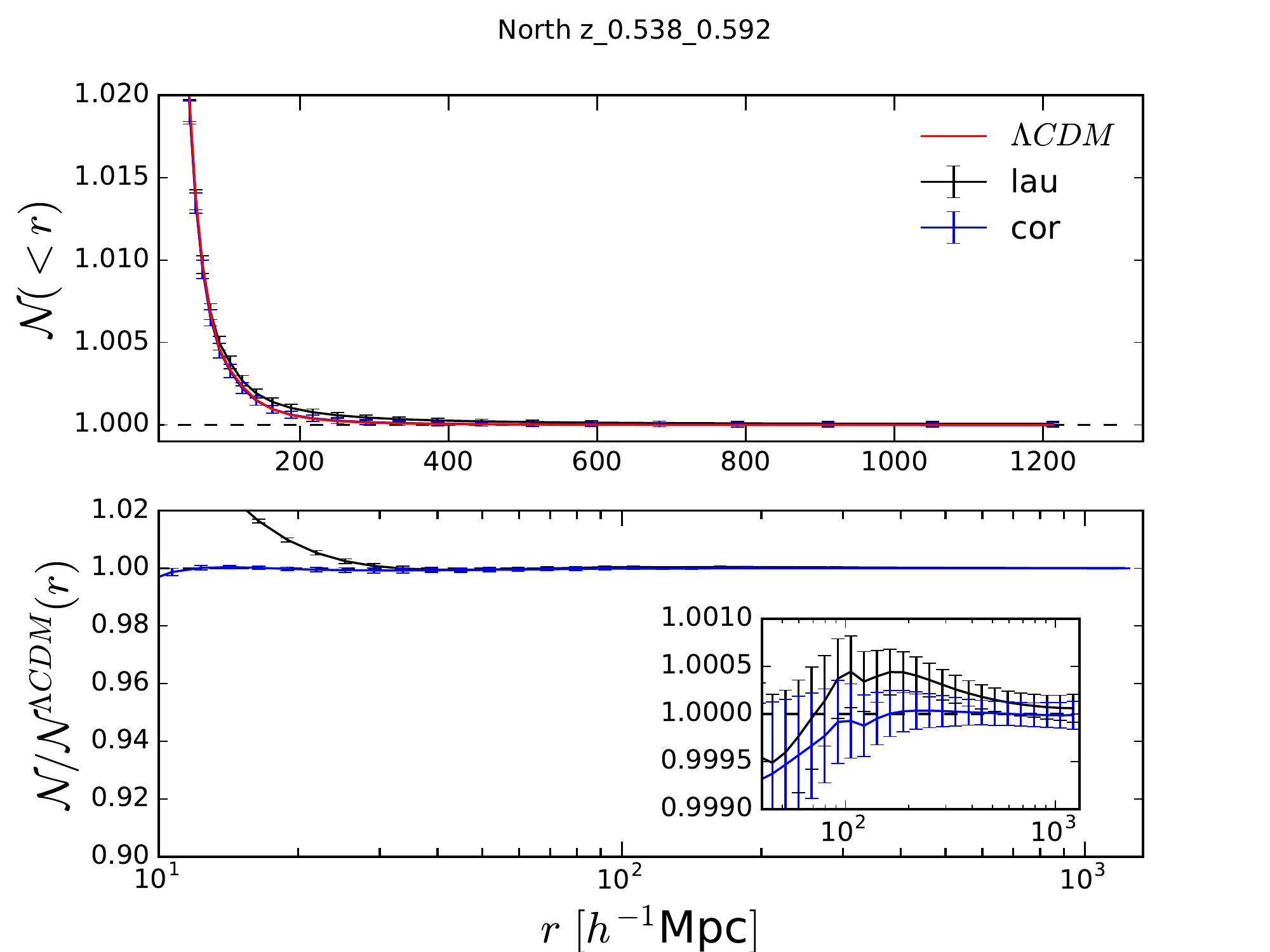}
	\includegraphics[width=0.48\linewidth, keepaspectratio]{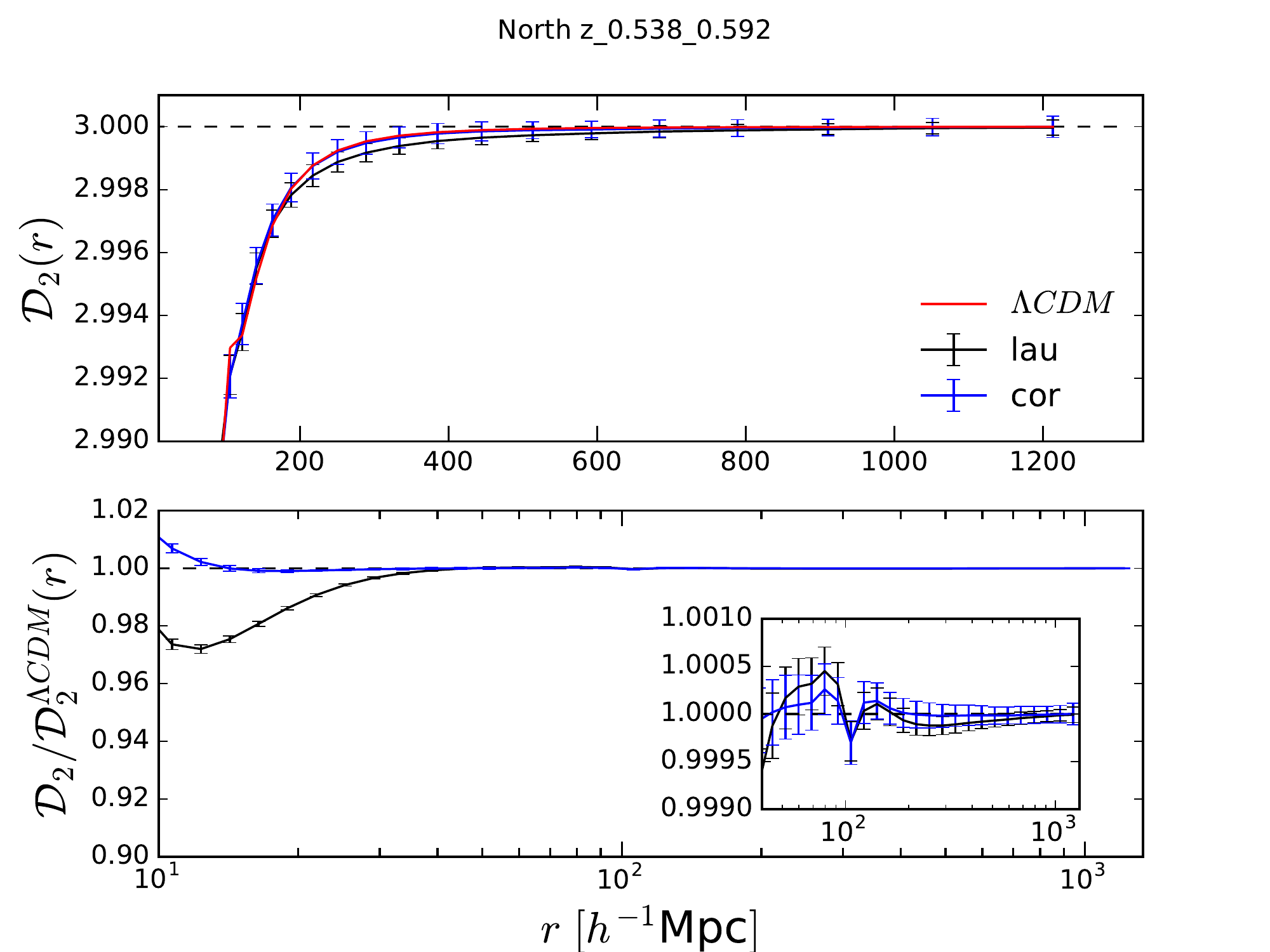}
	\caption{\label{fig:difference-Nest} 
Top: the scaled counts-in-spheres, $\mathcal{N}(<r)$, (left) and the fractal correlation dimension, $\mathcal{D}_{2}(r)$, (right) for matter distribution, with lau (black) and cor (blue) estimators, compared to $\Lambda$CDM model (red).
Bottom: the ratio to $\Lambda$CDM model for both estimator.}
\end{figure}

In section~\ref{subsect:Estimators}, we consider two estimators for ${\mathcal{N}}(<r)$, defined by equations (\ref{eq:lau}) and (\ref{eq:Counts}). We compute ${\mathcal{N}}(<r)$ and the resulting ${\mathcal{D}}_2(r)$ with the two estimators for the 1000 QPM mock catalogues. Figure \ref{fig:difference-Nest} compares the mean of the 1000 mocks to the $\Lambda$CDM model. The result with the cor estimator are much closer to the $\Lambda$CDM model. 

\section{Tuning errors in RSD Analysis}\label{APP:RSD-robust}

\begin{figure}[h!]
	\centering
	\includegraphics[width=.49\linewidth, keepaspectratio]{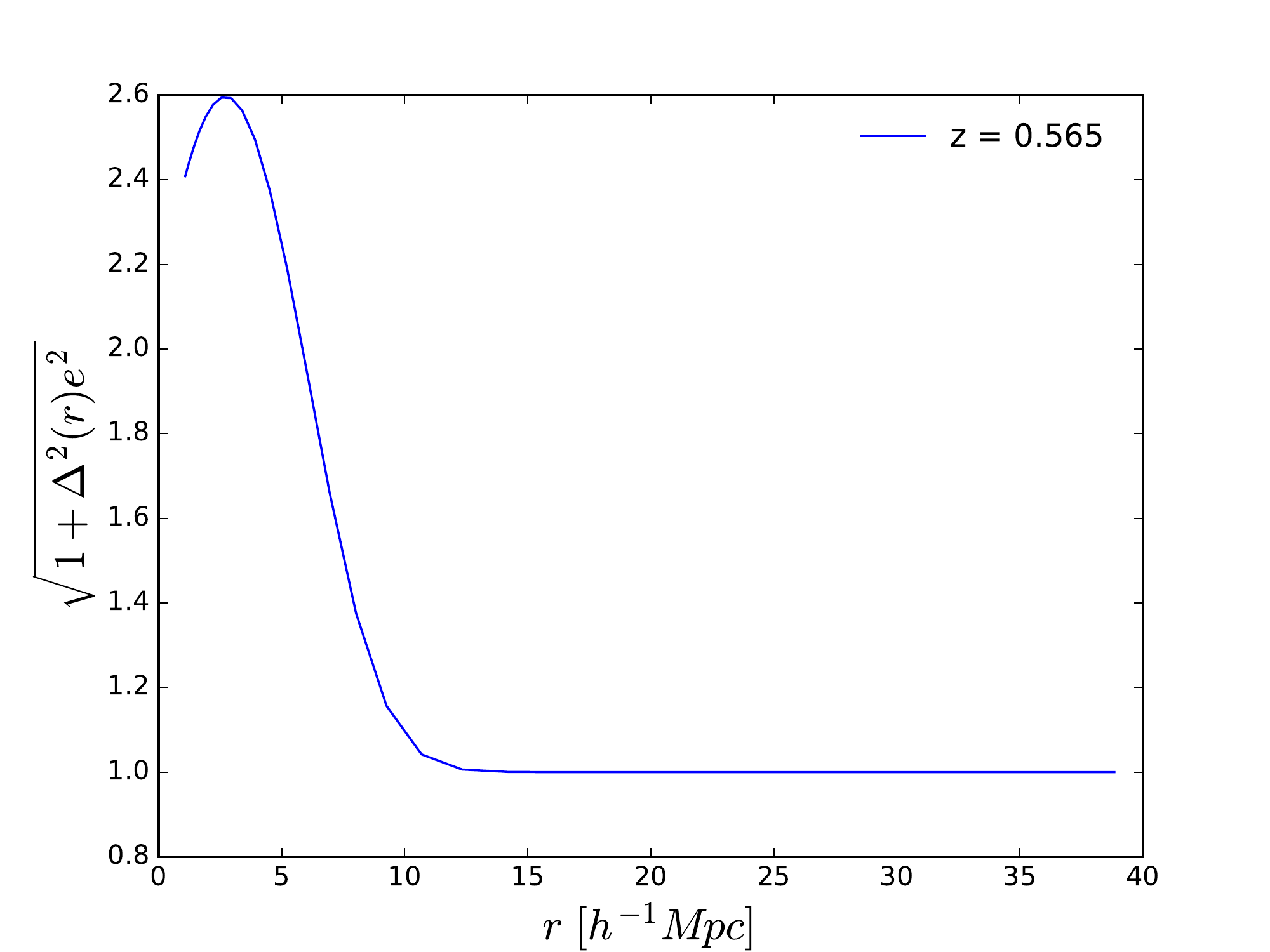}
	\includegraphics[width=.49\linewidth, keepaspectratio]{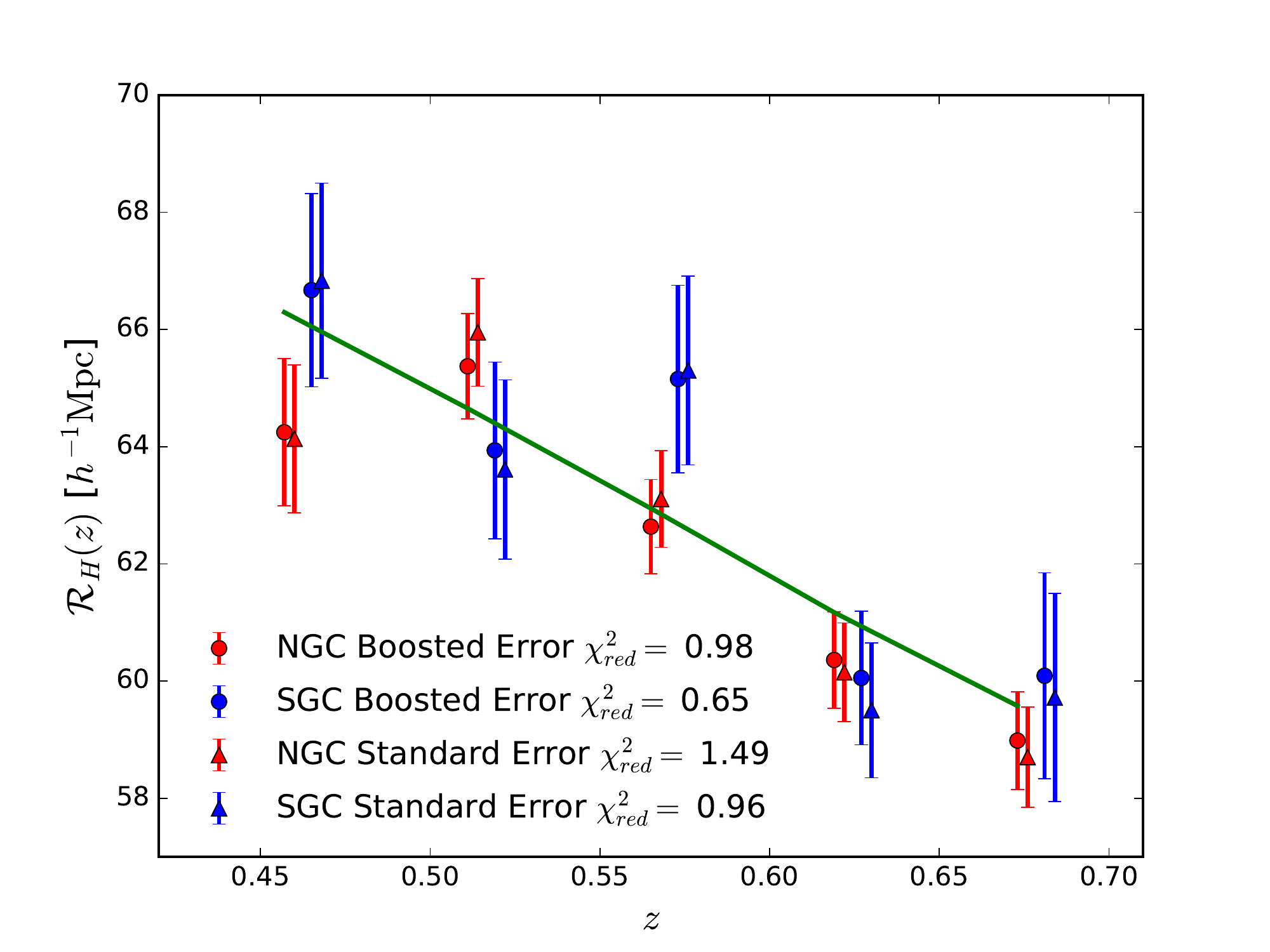}		
	\caption{\label{fig:Rh_z_effDiff} 
	Left: Boosting factor on the error in the correlation function. 
	Right: $R_H^{D_2=2.97}(z)$ with (circle) and without (triangle) boosting the errors, for NGC (red) and SGC (blue).}
\end{figure}

The theoretical model for redshift space distorsions (Eq.~\ref{eq:LSS_RSD}) is not perfectly accurate at the smallest scales due to the real nonlinear behaviour of gravity at these scales. In order to ensure satisfying $\chi^2/$n.d.f.\ for the RSD fitting,
we boost the error on $\xi(r)$ at the relevant scales in an empirical way as
\begin{equation}
		C^{'}_{ij} = C_{ij} ( 1 + \delta_{ij}\Delta_{i}\Delta_{j}e^2 ) \; .
\end{equation}
Here $\delta_{ij}$ is the usual Kronecker symbol; $e$ is a parameter that measures the amount of error boosting we apply; $\Delta_{i}$ is the theoretical inaccuracy, estimated as the relative difference between our model and the average of the $1000$ QPM mock catalogues. Figure \ref{fig:Rh_z_effDiff} (left) presents the resulting boosting factor on the error in the correlation function. It appears to be significant only on scales smaller than 10 $h^{-1}$ Mpc.
Fig.~\ref{fig:Rh_z_effDiff} (right) shows that the reconstructed homogeneity scales measured with $\mathcal D_2$ is not significantly modified by the error boosting.

\section{Test of spline fit on QPM mock catalogues}\label{subsec:Robustness}  	
 
We perform a spline fit of $\mathcal{D}_2$ for the 1000 QPM mock catalogues in order to obtain the homogeneity scale at 1\% (Eq. \ref{eq:RH-definition}). 
The fit is performed in the range $r \in [40,100]\ h^{-1}$ Mpc with $6$ data points and $1$ degree of freedom. The distribution of the $\chi^{2}$ of the mock should therefore follow a $\chi^{2}$-law for 1 degree of freedom. In table (\ref{tab:chi2Splines}) we show the mean and the error on the mean of the distribution of the corresponding $\chi^2$ for the $1000$ QPM mock catalogues. The test is successful in both NGC and SGC.

\begin{table}[h!]
		\begin{center} 
		\begin{tabular}{ *3c } 
	 	$z$   & $\chi^2_{\rm NGC}$ & $\chi^2_{\rm SGC}$  \\ 
	 	\hline 
		0.430-0.484 & $1.00\pm0.05$  & $0.99\pm0.04$   \\ 
	 	0.484-0.538 & $0.99\pm0.05$  & $1.00\pm0.04$   \\ 
	 	0.538-0.592 & $1.02\pm0.05$  & $1.00\pm0.04$   \\ 
		0.592-0.646 & $0.99\pm0.04$  & $1.00\pm0.05$   \\ 
		0.646-0.700 & $1.02\pm0.05$  & $1.00\pm0.04$   \\  
	 	\hline
		\end{tabular}
		\end{center} 
\caption{\label{tab:chi2Splines} Mean and error over the $1000$ QPM mock catalogues for the $\chi^2$ of the spline fit with 1 degree of freedom, in the NGC and SGC for the five redshift bins.}
\end{table}	
	
\section{Homogeneity scale at 0.1\%}

\begin{figure}[h!]
	\centering
	\includegraphics[width=100mm]{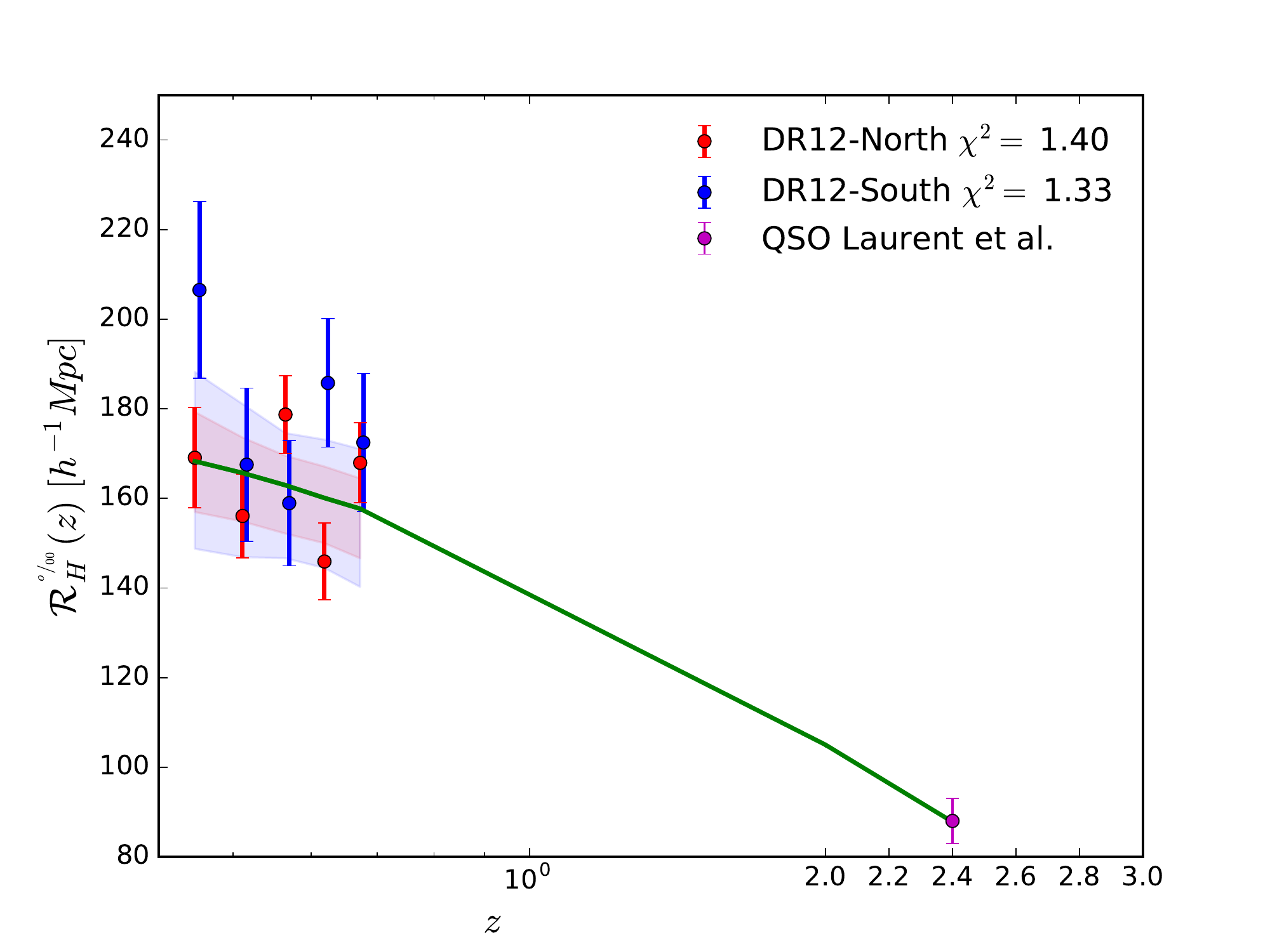}
	\caption{\label{fig:RH-z-perMil} The homogeneity scale at 0.1\% level, $R^{\mathcal{D}_2=2.997}_H(z)$, measured in the NGC (red) and in the SGC (blue) as a function of redshift. The purple point is the result obtained with quasars in the NGC,  in the redshift range $2.2\le z \le 2.8$ by \citet{Laurent}.
	The green line is the $\Lambda$CDM model prediction. The shaded areas indicate the 1$\sigma$ range for the 1000 QPM mock catalogues.}	
\end{figure}

The choice of a 1\% threshold to define the homogeneity scales is arbitrary. We can define them for instance at 0.1\% as:
	\begin{equation}\label{eq:RH-definition1}
		\mathcal{D}_{2}(R^{\mathcal{D}_2=2.997}_H) = 2.997 \quad {\rm or} \quad \mathcal{N}(R^{\mathcal{N}=1.001}_{H}) = 1.001
	\end{equation}
Figure \ref{fig:RH-z-perMil} shows that the measured homogeneity scale for matter distribution is compatible with $\Lambda$CDM, with 
$\chi_{red}^{2} =5.82/6$ in the NGC and $\chi_{red}^{2} = 7.98/6$ in the SGC.
	
\label{Bibliography}


\bibliographystyle{unsrtnat_arxiv} 

\bibliography{mybib} 

\end{document}